\documentclass[%
superscriptaddress,
amsmath,amssymb,
aps,
pra,
onecolumn,
]{revtex4-2}

\usepackage{amsfonts}
\usepackage{amsmath}
\usepackage{txfonts}
\usepackage{amssymb}
\usepackage{amsbsy}
\usepackage{graphicx}
\usepackage{dcolumn}
\usepackage{bm}
\usepackage{dsfont}
\usepackage{tikz}
\usetikzlibrary{quantikz}
\usepackage{booktabs,eqparbox}
\usepackage{xcolor}
\usepackage{braket}
\usepackage{algorithm}
\usepackage{algpseudocode}
\allowdisplaybreaks
\usepackage[breaklinks]{hyperref}
\hypersetup{colorlinks=true, linkcolor=blue, citecolor=blue, filecolor=blue, urlcolor=blue}
\graphicspath{{figures}}
\usepackage{mathtools}
\usepackage{tcolorbox} 
\newcommand{\tc}[1]{\textcolor{black}{#1}}

\begin{document}

\preprint{APS/123-QED}

\title{Resource-Efficient Bio-Molecular Docking on a NISQ-era Digital Quantum Computer}

\author{Tianqi Chen}
\email{chen\_tianqi@a-star.edu.sg}

 \affiliation{Bioinformatics Institute, Agency for Science, Technology and Research (A*STAR), 30 Biopolis Street, No.~07-01 Matrix, Singapore 138671\looseness=-1}
 \affiliation{A*STAR Quantum Innovation Centre (Q.~InC), Agency for Science, Technology and Research (A*STAR), 2 Fusionopolis Way, Innovis No.~08-03, Singapore 138634\looseness=-1}
 \affiliation{Institute of Advanced Intelligence and Computing (IAIC), Agency for Science, Technology and Research (A*STAR), 1 Fusionopolis Way, No.~16-16 Connexis, Singapore 138632\looseness=-1}
  \affiliation{Centre for Quantum Technologies, National University of Singapore, Singapore 117543}
  \affiliation{Department of Physics, National University of Singapore, Singapore 117551}

\author{Adrian M. Mak}
\email{makwk@a-star.edu.sg}
 \affiliation{Institute of Advanced Intelligence and Computing (IAIC), Agency for Science, Technology and Research (A*STAR),
1 Fusionopolis Way, No.~16-16 Connexis, Singapore 138632\looseness=-1}

\author{Jianguo Li}
\email{lijg@a-star.edu.sg}
\affiliation{Bioinformatics Institute, Agency for Science, Technology and Research (A*STAR), 30 Biopolis Street, No.~07-01 Matrix, Singapore 138671\looseness=-1}

\author{Jian Feng Kong}
\email{kongjf@a-star.edu.sg}
\affiliation{Institute of Advanced Intelligence and Computing (IAIC), Agency for Science, Technology and Research (A*STAR),
1 Fusionopolis Way, No.~16-16 Connexis, Singapore 138632\looseness=-1}
 \affiliation{A*STAR Quantum Innovation Centre (Q.~InC), Agency for Science, Technology and Research (A*STAR), 2 Fusionopolis Way, Innovis No.~08-03, Singapore 138634\looseness=-1}

\author{Chandra Verma}
\affiliation{Bioinformatics Institute, Agency for Science, Technology and Research (A*STAR), 30 Biopolis Street, No.~07-01 Matrix, Singapore 138671\looseness=-1}
\affiliation{Department of Biological Sciences, National University of Singapore, 14 Science Drive 4, Singapore 117543\looseness=-1}
\affiliation{School of Biological Sciences, Nanyang Technological
University, 60 Nanyang Drive, Singapore 637551\looseness=-1}
\author{Sebastian Maurer-Stroh}
\affiliation{Bioinformatics Institute, Agency for Science, Technology and Research (A*STAR), 30 Biopolis Street, No.~07-01 Matrix, Singapore 138671\looseness=-1}
\affiliation{Department of Biological Sciences, National University of Singapore, 14 Science Drive 4, Singapore 117543\looseness=-1}

\date{\today}

\begin{abstract}
Molecular docking is a vital computational task in drug discovery, wherein the objective is to efficiently identify optimal binding poses between a ligand and a target receptor protein. Due to the combinatorial explosion of possible binding configurations, docking of large and flexible molecules remains a computationally intensive problem, especially at scale. Early studies have revealed that the molecular docking can be re-cast as a maximum vertex-weighted clique problem (MVWCP) problem on a compatibility graph to be solved classically. In this work, we proposed a hybrid quantum-classical approach for molecular docking leveraging the MVWCP formalism with a variational full-basis encoding (FBE) strategy, which enables efficient encoding of classical binary variables with Bloch sphere vectors. \tc{We further prove that a global minimizer of the FBE objective can always be chosen to be a pure product state, thereby providing a rigorous justification for its optimization using a unitary variational circuit. }The molecular docking problem is first mapped to a cost Hamiltonian that is minimized within a variational framework, optimized via a randomized imaginary time evolution (ITE)-inspired warm start, and gradient-based techniques. Finally, we also executed the circuit on an IBM quantum computer, underlying the feasibility and utility of quantum-assisted optimization for structure-based drug design and point towards the broader utility of advanced encoding techniques in quantum optimization for computational biology.
\end{abstract}


\maketitle

\section{Introduction}

Recent years have witnessed rapid conceptual and technological advances in quantum technologies, and major efforts by industry and academia have led to programmable quantum processors exceeding one hundred qubits with steadily improving gate fidelities, establishing a versatile experimental platform for studying complex physical and computational problems~\cite{Arute2019supremacy,Jurcevic2021}. At the same time, landmark quantum algorithms such as Shor's factoring~\cite{shor1999polynomial} and Grover's search~\cite{grover1996fast} highlight the potential for quadratic speedups of unstructured search problems, but their practical realization requires fault-tolerant hardware operating far beyond current capabilities. Contemporary devices therefore operate in the Noisy Intermediate-Scale Quantum (NISQ) regime~\cite{Preskill2018quantumcomputingin}, where limited circuit depth, noise and qubit count motivate hybrid quantum--classical strategies in which classical high-performance computing performs substantial parts of the computational pipeline.

A central challenge for near-term quantum computing is the preparation of ground states of combinatorial Hamiltonians. Variational quantum algorithms (VQA)~\cite{Yuan_2019,Cerezo_2021} emerge and serve as a practical tool for these ground-state searching problems, and more recently classical combinatorial optimization problems such as the MAXCUT~\cite{Wang2018QAOA,koch2025quantumoptimizationbenchmarkinglibrary} or maximum independent set (MIS)~\cite{pichler2018quantumoptimizationmaximumindependent,ebadi2022quantum} can be solved via variational approaches like QAOA~\cite{Farhi2014QAOA} on Rydberg atom array-based quantum computers\cite{Saffman2010quantum,adams2020rydberg,wu2021concise,cong2022hardware,Shen2023proposal}. However, these are generally expected to require deep circuits for larger graph instances, which leads to significant hardware overhead and optimization instability, limiting their applicability on NISQ devices. This motivates the search for alternative formulations that reduce resource requirements while preserving the essential structure of the optimization problem.

Molecular docking provides a natural and scientifically important setting for such combinatorial optimization~\cite{Kitchen2004Docking,Meng2011Docking,Paggi2024DockingReview,Agu2023Docking}.
It aims to identify the most probable binding configuration between a ligand and a target protein, a task central to structure-based drug discovery~\cite{Kitchen2004Docking,Meng2011Docking,Shirali2025Scoring,Paggi2024DockingReview}. Despite decades of algorithmic progress, it remains demanding for docking large and flexible molecules because of the inefficient sampling of the vast ligand-receptor configurational space. The number of candidate poses grows combinatorially with ligand flexibility, interaction motifs and binding-site geometry, making exhaustive evaluation inefficient. In practice, classical pipelines rely on heuristic search, scoring approximations and aggressive pruning, which can struggle in regimes where interaction constraints are dense, correlated or geometrically ambiguous.

Therefore, a seemingly natural approach to this difficulty is to reformulate docking as a graph-based combinatorial problem~\cite{kuhl1984combinatorial,Ding2024DockingQAOA,papalitsas2025quantumapproximateoptimizationalgorithms}. By representing pharmacophore interactions as vertices in a compatibility graph, feasible docking poses correspond to cliques, and the optimal binding configuration is obtained by solving the maximum vertex-weighted clique problem. This mapping separates geometric feasibility from interaction scoring while exposing a hard discrete optimization problem. Recent works~\cite{Ding2024DockingQAOA,papalitsas2025quantumapproximateoptimizationalgorithms} which are based on this formalism showcase the successful identification of target pharmacophore points, yet require heavy computational overhead and resources. In addition, we also remark there are other recent works that share similar scope on other quantum platforms with different classical encoding formalism~\cite{Banchi_2020,yu2023universal,Wen2026}, \tc{as well as using a quantum circuit evolution (QCE) method by applying random unitaries without the calculation of gradients~\cite{dejesus2026moleculardockingquantumcircuit}.}

Here we introduce a resource-efficient variational quantum framework for maximum vertex-weighted clique problem (MVWCP)-based molecular docking built on multi-basis encoding~\cite{Patti2022MBE}. The central idea is to represent multiple classical variables using single-qubit observables across different measurement bases, enabling qubit compression while introducing geometric coupling through Bloch-sphere constraints. \tc{Importantly, we prove that at least one global minimizer of the resulting FBE objective can always be represented by a pure product state, establishing that entanglement is not required to represent an optimal solution and providing a rigorous foundation for its optimization using a unitary variational circuit.} Combined with a warm-start strategy inspired by imaginary-time evolution, this approach reshapes the optimization landscape so that correlated interaction patterns can be explored using comparatively shallow circuits. Rather than replacing classical docking pipelines, the method targets the combinatorial subproblem of selecting mutually compatible interaction sets, for which compact correlated representations may offer a practical advantage on near-term quantum devices.

The remainder of this paper is organized as follows. 
In Section~\ref{sec:methods}, we first formulate the generic QUBO/Ising 
optimization problem and develop the warm-start full-basis encoding 
(W-S FBE) framework, including the Bloch-vector encoding and the stochastic 
imaginary-time-evolution-inspired initialization, before applying it to 
MVWCP-based molecular docking. In Section~\ref{sec:results}, we construct two 
biologically relevant docking instances and present the numerical and hardware 
results, including optimization convergence and clique recovery, execution on 
an IBM superconducting quantum processor, comparison with the ZX-basis 
encoding, and sensitivity to circuit depth and penalty strength. 
In Section~\ref{sec:conclusions}, we summarize the main findings, discuss their 
implications and limitations, and outline directions for future research. 
The Appendices provide a proof that a global minimizer of the FBE objective can 
be chosen as a pure product state, together with further details of the hardware 
implementation, docking graphs, evaluation metrics, and the complete W-S FBE 
algorithm.

\section{Models and Methods}
\label{sec:methods}

\begin{figure}[t]
    \centering
    \includegraphics[width=1.0\columnwidth,draft=false]{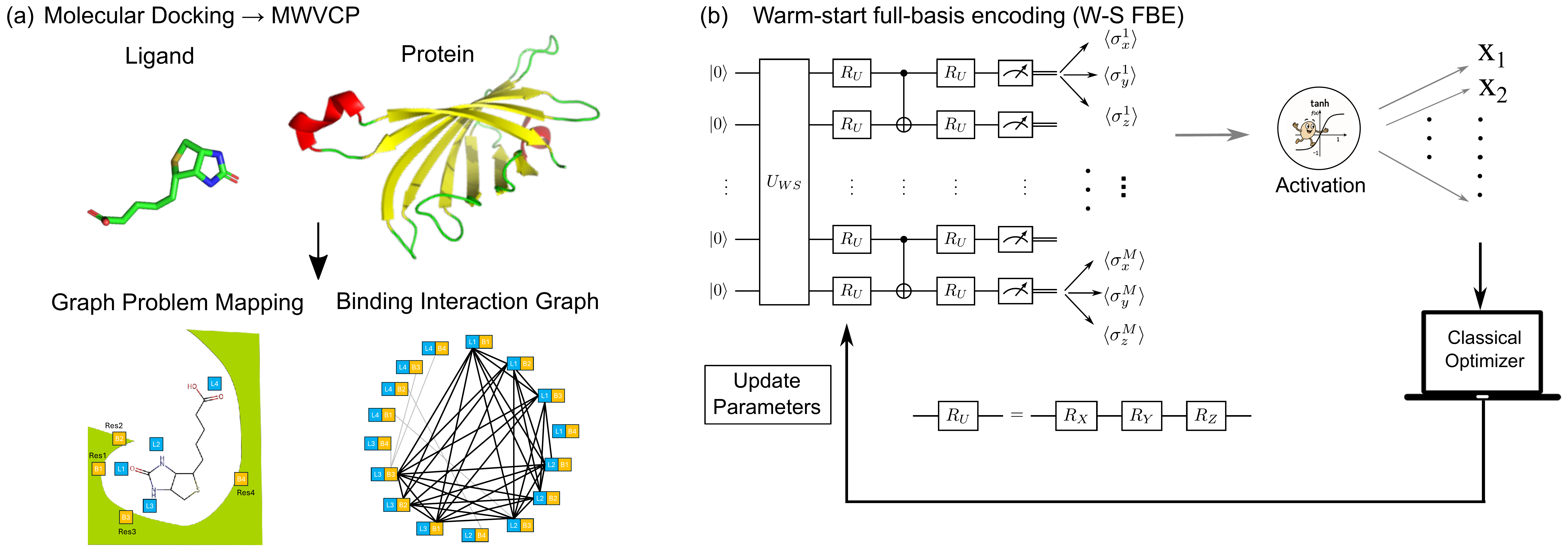}
    \caption{{Illustration of the problem formalism and multi-basis encoding approach: \textcolor{black}{(a) The molecular docking of ligand and protein is mapped to finding a maximum vertex-weighted clique on a graph: selected pharmacophores from both ligand (blue) and protein (orange) are analyzed and mapped to a labeled distance graph (LDG), following the procedure and criteria introduced in Ref.~\cite{Ding2024DockingQAOA}. The binding interaction graph (BIG) is then defined based on the LDG, with its vertices representing the product of the selected pharmacophores from ligand and protein. Here, we used \texttt{1stp} as an example for illustration.} (b) Full-basis encoding (FBE) with an initial warm-start layer \textcolor{black}{$U_{WS}$}: for each qubit from a parametrized variational quantum circuit, all the three axes in the Bloch sphere ($\langle \sigma_x \rangle$, $\langle \sigma_y\rangle$, $\langle \sigma_z \rangle$) are measured, followed by an hyperbolic activation function \textit{$\tanh$} such that these activated observables are substituted for each classical variable $x_i$ from the combinatorial problem. Here, the tuple of measured observables correspond to: $(\langle \sigma_z^i \rangle, \langle \sigma_x^i \rangle, \langle \sigma_y^i \rangle) \rightarrow (x_{3i-2}, x_{3i-1}, x_{3i})$ ($i=1,2,\cdots,M$, where $M$ is the number of qubits for the circuit). The cost function is further optimized to obtain the minimum value on a classical computer, and concurrently update the parameters in the quantum circuit. For an $N$-sized combinatorial optimization problem, only $M=\lceil N/3 \rceil$~\cite{roundupnote} qubits are used. Here, an example of $M=4$ circuit is shown here for the purpose of illustration. The parameters for the gates within the warm-start layer $U_{WS}$ is obtained from the results given by the warm-start via imaginary time evolution~\cite{ChaiWarmStart2025} (see Sec.~\ref{sec: warm-start} for more details). Finally, the optimal clique (filled magenta circles) and its corresponding docking contacts are obtained once an optimal bitstring is obtained from the optimization. }}
    \label{fig: illustration}
\end{figure}
\footnotetext{Here, the symbol $\lceil \cdots \rceil$ represents the smallest integer greater than or equal to the input. As a result,if $M$ is not divisible by $3$, then for the last qubit, some observables are not utilized to substitute the classical variable from the combinatorial problem.}

\subsection{\tc{Warm-Start Full-Basis Encoding (W-S FBE)}}
\label{sec:warm-start FBE}

\subsubsection{The Quadratic Unconstrained Binary Optimization problem and its transformation}

The quadratic unconstrained binary optimization (QUBO) problem offers a unifying framework for combinatorial optimization: its objective couples binary variables through linear and pairwise quadratic terms alone, yet a broad class of NP-hard problems---graph partitioning, maximum cut, set cover, and maximum-weight clique among them---can be cast in this form by folding their constraints into quadratic penalty terms~\cite{lucas2014ising, kochenberger2014unconstrained}. Because a QUBO objective maps directly onto the Ising Hamiltonian of statistical physics under a simple change of variables, it has become the de facto input format for quantum and quantum-inspired optimizers, from quantum annealers such as D-Wave systems~\cite{johnson2011dwave} to the gate-based Quantum Approximate Optimization Algorithm (QAOA)~\cite{glover2019tutorial, Farhi2014QAOA, date2021qubo}. We set out this formalism below, as it underlies the molecular-docking instances treated in Sec.~\ref{sec:docking MVWCP}.

Consider $N$ binary decision variables $x_j\in\{0,1\}$ ($j=1,\dots,N$), together with linear weights $w_j$ and pairwise couplings $J_{ij}$. The associated quadratic unconstrained binary optimization (QUBO) problem reads

\begin{align}
    \label{eq: generic QUBO}
    &\text{Maximize}\,\,\,\, \mathcal{C}_0=\sum_{i \neq j, i,j=1}^N J_{ij}x_i x_j + \sum_{j=1}^N w_j x_j, \\ \nonumber
    &\text{Subject to}\,\,\,\, x_j \in \{0,1\}.
\end{align}

where $J_{ij}$ and $w_j$ defines the pairwise weight for $(i,j)$ and the linear weight for the binary variable $j$, respectively. Maximizing $\mathcal{C}_0$ favours assignments that respect the underlying problem constraints in this paper. To bring this into a form amenable to quantum optimization, we map the binary variables onto spin-$1/2$ variables $z_j\in\{-1,1\}$ via the transformation $x_j=\left(z_j+1\right)/2$ and add an overall minus sign, which turns the maximization into a minimization, posing it as a ground-state searching problem.

\begin{align}
    \label{eq: generic Ising}
   &\text{Minimize}\,\,\,\, \mathcal{C}_1=\sum_{i \neq j, i,j=1}^N -\frac{J_{ij}}{4}\left(z_i z_j + z_j + z_i+1\right) - \sum_{j=1}^N \frac{1}{2}w_j \left(z_j+1\right), \\ \nonumber
    &\text{Subject to}\,\,\,\, z_j \in \{-1,1\}.
\end{align}

Finally, promoting each spin variable $z_j$ to a Pauli-$Z$ operator $\hat{\sigma}_j^z$ and the scalar $1$ to the identity operator $\hat{I}$ \textcolor{black}{(which can be further omitted as it only contributes to an energy shift of the total discrete energy spectrum of the Hamiltonian)} recasts Eq.~\eqref{eq: generic Ising} as the task of finding the ground state of an Ising-type Hamiltonian.

\begin{align}
    \label{eq:Pauli-Z Ising}
    &\hat{H}_{\text{Pauli-Z}}=\sum_j \beta_j \hat{h}_j \\ \nonumber
    &=\sum_{i \neq j, i,j=1}^N -\frac{J_{ij}}{4}\left(\sigma^z_i \sigma^z_j + \sigma^z_j + \sigma^z_i\right) - \sum_{j=1}^N \frac{1}{2}w_j \sigma^z_j
\end{align}

where we have re-written it in the generic Pauli form $\hat{H}=\sum_{j}\beta_j\hat{h}_j$, where each $\hat{h}_j$ is a Pauli string (a single $\hat{\sigma}^z_i$ or a product $\hat{\sigma}^z_i\hat{\sigma}^z_{i'}$) and $\beta_j$ its coefficient. Any problem expressible in the QUBO form of Eq.~\eqref{eq: generic QUBO}---including the molecular-docking instances constructed later in Sec.~\ref{sec:docking MVWCP}---can therefore be solved by the encoding procedure described next. 

\subsubsection{From Multi-Basis Encoding (MBE) to Full-Basis Encoding (FBE): hosting several variables on a single qubit}

Here, we introduce an encoding procedure designed to minimize a generic quadratic cost function over binary variables in the QUBO form, independently of where that cost function comes from. In the most common variational treatment of an Ising problem such as Eq.~\eqref{eq: generic Ising}, a separate qubit is assigned to each binary variable: the $j$-th variable is read out from the single observable $\langle\sigma_j^z\rangle$, so that an $N$-variable problem requires $N$ qubits. On NISQ hardware, where qubit count, circuit depth and coherence time are all severely limited, this one-variable-per-qubit mapping is wasteful and quickly becomes a bottleneck.

The starting point for a more economical encoding is the observation that a single qubit already carries \emph{three} independent pieces of information. The state of one qubit is fully specified by its Bloch vector, whose components are the three expectation values $\langle\sigma^x\rangle,\langle\sigma^y\rangle,\langle\sigma^z\rangle$, each lying in the interval $[-1,1]$. Since the three Bloch axes are independent, each one can be used to carry the value of a \emph{different} classical variable. This is the central idea of the multi-basis encoding (MBE) of Ref.~\cite{Patti2022MBE}: rather than encoding one variable per qubit, MBE hosts two variables per qubit by reading them out from two complementary measurement bases, $\langle\sigma^z\rangle$ and $\langle\sigma^x\rangle$. Concretely, each measured expectation value $\langle\sigma^\zeta\rangle\in[-1,1]$ is passed through a smooth, bounded activation function---here the hyperbolic tangent, $\Phi(\cdot)=\tanh(\cdot)$---to produce a continuous, ``relaxed'' surrogate for a binary variable. The original combinatorial cost function is then evaluated on these relaxed variables, and a definite binary assignment is recovered at the end of the optimization simply by taking the sign of the corresponding expectation value. The activation plays a role analogous to a nonlinearity in a classical neural network: it keeps the relaxed variables bounded and regularizes the joint optimization of the several variables that now share a single qubit. In this way MBE halves the qubit requirement relative to the naive mapping while keeping the circuit shallow, which is precisely what makes it attractive for near-term quantum devices.

We now push this idea to its natural limit. Because a qubit has three independent Bloch components, \emph{all three} Pauli observables---and not just the two of MBE---can be used to host classical variables, so that up to three variables are encoded per qubit. In this work, we generalize multi-basis encoding (MBE) to a full-basis encoding (FBE) that exploits all three Pauli observables on each qubit. Let $\mathcal{G}=(V,E)$ be a weighted graph with the total number of vertex $|V|=N$, the total number of edges $E$. We introduce an embedding $\Pi$ as

\begin{align}
\Pi:V\rightarrow \{(q,\zeta): q\in\{1,\dots,m\},~\zeta\in\mathcal{B}\},
\end{align}

which assigns each vertex to a qubit index $q$ and a measurement basis $\zeta$ which belongs to $\mathcal{B}=\{x,y,z\}$. Because each qubit supports three orthogonal observables, up to three vertices can be hosted per qubit, so that $m=\lceil N/3\rceil$ qubits suffice \tc{($\lceil \cdots\rceil$ stands for rounding a number up to the nearest whole integer number)}. Now, given a parametrized quantum state with shallow circuits as in Fig.~\ref{fig: illustration}(b):

\begin{equation}
|\psi(\mathbf{\theta})\rangle=U(\mathbf{\theta})|0\rangle^{\otimes m},
\end{equation}

we define single-qubit expectation values

\begin{equation}
s_q^{\zeta}(\mathbf{\theta})
=
\langle \psi(\mathbf{\theta})|\sigma_q^{\zeta}|\psi(\mathbf{\theta})\rangle .
\end{equation}

Here, $\mathbf{\theta}$ are the parameters for the rotation gates ($R_x$, $R_y$ and $R_z$) in each layer [Fig.~\ref{fig: illustration}(b)]. Each vertex $v$ is represented by a relaxed variable

\begin{equation}
x_v(\mathbf{\theta})
=
\Phi\!\left(
s_{q(v)}^{\zeta(v)}(\mathbf{\theta})
\right),
\end{equation}

where $\Phi(\cdot)$ is a bounded nonlinear activation function evaluated classically. In this work, we have chosen the hyperbolic tangent function $\tanh$ [Fig.~\ref{fig: illustration}(b)]. The activation function regularizes simultaneous optimization of multiple observables hosted by the same qubit. In this regard, the FBE objective is constructed from products of single-qubit expectation values rather than direct multi-axis operator expectations. For a problem in the QUBO form of Eq.~\eqref{eq: generic QUBO}, this yields the (raw) cost function to be minimized:
\begin{align}
&\mathcal{L}_{\mathrm{FBE}}^{\mathrm{raw}}(\mathbf{\theta})
=-\left[\sum_{u \neq v, u,v=1}^N J_{uv}x_u(\mathbf{\theta}) x_v(\mathbf{\theta}) + \sum_{u=1}^N w_u x_u(\mathbf{\theta})\right]
\label{eq:LFBE}
\end{align}

Again, $J_{uv}$ defines the pairwise weight for $(u,v)$ and $w_u$ represents the linear weight for the binary variable at site $u$, respectively. This cost function $\mathcal{L}_{\mathrm{FBE}}^{\mathrm{raw}}(\mathbf{\theta})$ in Eq.~\eqref{eq:LFBE} is defined as the \textit{raw cost function}, which is then optimized classically. It can be further proved that under the scheme of full-basis encoding, the ground state of Eq.~\eqref{eq:LFBE} can be exactly represented as a product state of an $m$-qubit quantum circuit, providing solid foundation for the usage of unitary quantum circuit optimization. The details of this proof is summarized in the Appendix.~\ref{sec:proofofgroundstate}. The entire process is summarized in Fig.~\ref{fig: illustration}(b).

Here, we also remark that this whole formalism reduces to two-basis MBE~\cite{Patti2022MBE} when only $\{x,z\}$ are used. The embedding introduces geometric constraints arising from the Bloch sphere. For each qubit, $\left(s_q^x\right)^2+\left(s_q^y\right)^2+\left(s_q^z\right)^2 \le 1$, which couples the variables hosted on the same qubit. Optimization therefore becomes a multi-axis constrained relaxation in which improving one encoded variable restricts the attainable values of the others. This provides an intrinsic regularization mechanism and biases training toward jointly stable assignments across bases. In addition, the quantum resources required to evaluate Eq.~\eqref{eq:LFBE} are the single-qubit expectation values,
so measurement cost scales as \textcolor{black}{$\mathcal{O}(m|\mathcal{B}|)$.}

After performing the gradient-descent optimization with Adam algorithm~\cite{kingma2017adammethodstochasticoptimization} on the raw cost function $\mathcal{L}_{\mathrm{FBE}}^{\mathrm{raw}}(\mathbf{\theta})$, a discrete assignment is obtained via rounding

\begin{align}
\breve{x}_v(\mathbf{\theta}_\star)
=
\mathrm{sign}\!\left(
s_{q(v)}^{\zeta(v)}(\mathbf{\theta}_\star)
\right),
\end{align}

where the rounding is used only for evaluation and does not affect gradient updates. Here, $\mathbf{\theta}_\star$ is the optimized circuit parameters. We can therefore, \textcolor{black}{in the optimization process}, define

\begin{align}
    \label{eq: rounded cost function}
    &\mathcal{L}_{\mathrm{FBE}}^{\mathrm{round}}(\mathbf{\theta})=\sum_{u \neq v, u,v=1}^N J_{uv}\breve{x}_u(\mathbf{\theta}) \breve{x}_v(\mathbf{\theta}) + \sum_{u=1}^N w_u \breve{x}_u(\mathbf{\theta})
\end{align}

as the \textit{rounded cost function} \textcolor{black}{to be minimised} as one of the performance metrics apart from the aforementioned raw cost function in Eq.~\eqref{eq:LFBE}. Finally, the corresponding transient clique weight value can be written as

\begin{align}
&\mathcal{L}_{\mathrm{FBE}}^{\mathrm{transient}}(\mathbf{\theta}_\star;G)
=
\frac{1}{2}
\sum_{(u,v)\in E}
w_{u}\breve{x}_u + w_{v}\breve{x}_v
.
\label{eq:cutFBE}
\end{align}
which is another performance metric to be minimised. This quantity is also monitored during the FBE optimization.

\subsubsection{Warm Start Inspired by Trotterized Stochastic Drift Imaginary Time Evolution}
\label{sec: warm-start}

Being a variational quantum algorithm to obtain the ground state, FBE requires not only many random starts which require a different initial state preparation for each time, but also iterative steps during the feed-forward optimization, where significant computational resources are needed. To circumvent these computational overheads, make the process resource-efficient, and enhance the performance of our full-basis encoding method on a NISQ-era quantum processor, we designed a warm-start approach inspired by quantum stochastic drift protocol (qDRIFT)~\cite{Campbell2019qDRIFT}. qDRIFT is a randomized product-formula protocol for Hamiltonian simulation in which, rather than applying every term of $\hat{H}$ in a fixed Trotter sequence, each elementary step is a single term $\hat{h}_j$ drawn at random with probability proportional to its weight $|\beta_j|$ and evolved for a uniform time increment. Its defining feature is that the number of gates required is controlled by the sum of the coefficient magnitudes $\lambda=\sum_j|\beta_j|$ instead of the number of Hamiltonian terms, which makes it attractive for Hamiltonians composed of many Pauli strings. We adopt this stochastic term-sampling idea, but repurpose it in two essential ways described below.

To begin, the imaginary time evolution of a Hamiltonian $\hat{H}$ typically guarantees convergence to its ground state $|{\psi_{\text{GS}}}\rangle$:

\begin{align}
    \label{eq: ITE definition}
    &|{\psi_{\text{GS}}}\rangle=\lim_{t\rightarrow\infty}\frac{e^{-t\hat{H}}|{\psi_0}\rangle}{\sqrt{\langle\psi_0|e^{-2t\hat{H}}|\psi_0\rangle}},
\end{align}

as long as the initial state $|\psi_0\rangle$ has some overlap with the ground state $|{\psi_{\text{GS}}}\rangle$. However, the imaginary time evolution operator $e^{-t\hat{H}}$ is non-unitary, making it difficult to be implemented even on a perfect quantum computer as its decomposition to fundamental quantum gates is non-trivial. Recently, there have been works implementing quantum imaginary time evolution with parametrized variational circuits~\cite{Liu2021probabilistic,Lin2021,kosugi2022imaginary,nishi2023optimal,turro2022imaginary,turro2023quantumimaginarytimepropagation,chen2023high,chen2023robust,chen2024efficient,leadbeater2024non,ejima2025probabilistic,shen2025robust,ma2025penaltyfreequantumalgorithmenergy,shen2025observation}. These existing approaches, however, commonly demand substantial resources to achieve precise parameter estimation. Inspired by a recent warm-start variational quantum algorithm for QUBO problems~\cite{ChaiWarmStart2025}, we perform a warm start by a few steps of initial imaginary time evolution of $\hat{H}_{\text{Pauli-Z}}$ of Eq.~\eqref{eq:Pauli-Z Ising} classically with matrix product states (MPS), i.e., 
\begin{align}
    \label{eq: ITE initial interation}
    &|{\psi_{\text{WS}}}\rangle=\frac{\prod_{j=1}^{\mathcal{N}_{\text{T}}}e^{-\Delta t_{\text{WS}}\hat{h}_j}|{\psi_{+}}\rangle}{\sqrt{\langle\psi_{+}|e^{-2\Delta t_{\text{WS}}\hat{h}_j}|\psi_{+}\rangle}},
\end{align}

where $|\psi_{+}\rangle=\otimes_{k=1}^N|+\rangle$ is an initial product state of $|+\rangle=1/\sqrt{2}\left(|0\rangle+|1\rangle\right)$, where $N$ is the total system size from the Hamiltonian $\hat{H}_{\text{Pauli-Z}}$. $\mathcal{N}_{\text{T}}$ is the number of Trotterized steps~\footnote{Throughout this work, we find that by having roughly Trotterized steps being between $10$ to $15$ steps will guarantee a good performance of FBE.}, and $\Delta t_{\text{WS}}$ is the time step. $\hat{h}_j$ is a term randomly drawn from the total Hamiltonian $\hat{H}_{\text{Pauli-Z}}=\sum_j \beta_j\hat{h}_j$ according to its coefficient probability distribution defined as

\begin{align}
    \label{eq: MC probability of coefficient}
    &{\mathbf p}(\hat{h}_j)
    =\frac{\left|\beta_j\right|}{\sum_{j=1}^\mathcal{M}\left|\beta_j\right|}.
\end{align}

Although Eq.~\eqref{eq: MC probability of coefficient} inherits the coefficient-weighted sampling of qDRIFT, our construction departs from it in two respects. First, whereas qDRIFT compiles \emph{real-time}, unitary dynamics $e^{-it\hat{H}}$ intended to run on a quantum processor, we apply the same stochastic sampling to \emph{imaginary-time} evolution $e^{-\Delta t_{\text{WS}}\hat{h}_j}$, whose non-unitarity drives the state toward the ground state rather than propagating dynamics. Second, we execute these drift steps \emph{classically} on a matrix-product-state backend from Qiskit~\cite{Qiskit2024}, and use the resulting state solely to warm-start the downstream FBE optimization, instead of realizing the evolution as a quantum circuit. This repurposing retains the Monte-Carlo character of qDRIFT---so that the residual uncertainty is purely statistical in origin---while sidestepping the non-unitarity that would otherwise obstruct a direct gate-level implementation.

After a sufficient number of time steps, the Pauli observables for each physical site $i$ ($i=1,2,\cdots,N$) is measured for all three basis: $\langle\sigma_i^\alpha\rangle$ ($\alpha=x,y,z$). We then supply a warm-start layer consisting of an $R_x(\phi_i)$, an $R_y(\theta_i)$ and an $R_z(\zeta_i)$ gate using the measured outcomes [see Fig.~\ref{fig: illustration}(b)] as the new initial state for the subsequent FBE optimization: $\ket{\psi_{\text{ini}}}=\bigotimes_{i=1}^M\left[R_x(\phi_i)R_y(\theta_i)R_z(\zeta_i)\ket{0}_i\right]$. The parameters of the three rotation gates are calculated as follows:

\begin{align}
    &\theta_i=\arcsin{\left\langle\sigma_i^x\right\rangle}, \\ \nonumber
    &\phi_i={\rm atan2}\left(-\langle\sigma_i^y\rangle,\langle\sigma_i^x\rangle \right).
\end{align}

where $\rm{atan2}$ returns the angle between the positive $x$-axis and the vector from the origin to the point $(x,y)$, with the correct quadrant taken into account. We remark that $\zeta_i$ can be chosen arbitrarily as it just contributes to an additional phase to the quantum state and does not affect $\langle \sigma_i^z\rangle$. The above whole process will ensure that after a few Trotterized steps, the final state, which is also the initial state supplied for FBE, has more degrees of overlap with the ground state than the initial product state $|\psi_{+}\rangle=\otimes_{k=1}^N|+\rangle$, saving the resource for the later-stage FBE optimization. Also, the stochastic nature of the procedure ensures that the resulting uncertainty is purely of Monte Carlo origin. Increasing the number of iterations systematically improves the approximation, yielding states that approach the ground state with higher fidelity. This hybrid strategy performs this stage classically, thereby facilitating the subsequent FBE stage.

Finally, we remark that the W-S ITE algorithm could possibly be done either by performing a mid-circuit measurement~\cite{ma2025penaltyfreequantumalgorithmenergy} or via a recently developed mimicking-ITE~\cite{chai2025optimizingquboquantumcomputer}. We have summarized the full algorithm of W-S FBE in Algorithm.~\ref{alg:warm_start_fbe} in the Appendix.

\subsection{Application: Molecular Docking as a Maximum Vertex-Weighted Clique Problem (MVWCP)}
\label{sec:docking MVWCP}

\tc{We now describe the docking problem which the W-S FBE scheme above is applied to, and show that it is exactly an instance of the QUBO problem of Eqs.~\eqref{eq: generic QUBO} to \eqref{eq: generic Ising}. Molecular docking seeks the most probable binding configuration between a ligand and a target protein \cite{Kitchen2004Docking,Meng2011Docking,Paggi2024DockingReview}. Direct exploration of the continuous configurational space is computationally demanding due to the large number of translational, rotational, and conformational degrees of freedom \cite{Kitchen2004Docking,Paggi2024DockingReview}. A common strategy is therefore to reformulate docking as a discrete combinatorial optimization problem by abstracting molecular geometry into pharmacophore representations and encoding interactions using graphs \cite{kuhl1984combinatorial,Banchi_2020,Ding2024DockingQAOA}.
}

\tc{Both the ligand and the protein binding site are reduced to sets of pharmacophore points that capture key interaction features, including charged groups, hydrogen-bond donors or acceptors, hydrophobic regions, and aromatic rings \cite{Kitchen2004Docking,Shirali2025Scoring,Banchi_2020}. These points define the vertices of a labeled distance graph (LDG), in which vertex labels represent pharmacophore types and edges encode pairwise Euclidean distances \cite{Banchi_2020,Ding2024DockingQAOA}. This representation preserves the geometric arrangement of interaction features while significantly reducing the dimensionality of the docking problem.
}\tc{
Given the LDGs for the ligand and the protein, possible intermolecular contacts are modeled by constructing a binding interaction graph (BIG) \cite{Banchi_2020,Ding2024DockingQAOA}. Each BIG vertex represents a potential contact between a ligand pharmacophore and a protein pharmacophore. If the ligand contains $n$ pharmacophore points and the protein contains $m$ points, the BIG therefore contains $n\times m$ vertices corresponding to all possible contact assignments.
}\tc{
Edges in the BIG encode compatibility between contacts. Compatibility is typically determined by geometric consistency: the distance between two pharmacophores on the ligand should approximately match the corresponding distance on the protein within tolerance parameters that account for molecular flexibility and interaction-specific distance uncertainty. Thus, two contact vertices are connected when they can coexist within a physically feasible docking pose \cite{kuhl1984combinatorial,Banchi_2020}.  In this way, the BIG enforces structural constraints while allowing limited flexibility.
}

To capture heterogeneous interaction strengths, each BIG vertex is assigned a weight reflecting the interaction potential of the corresponding pharmacophore pair \cite{Kitchen2004Docking,Shirali2025Scoring}. For instance, a pharmacophore pair consisting of a hydrogen bond donor and a hydrogen bond acceptor would yield a larger weight, whereas a pair consisting two pharmacophores having local negative charges would yield a lower weight. These weights may be derived from empirical contact potentials, statistical models, or physics-based interaction estimates. Consequently, the BIG simultaneously encodes geometric feasibility through its edges and interaction strength through its vertex weights.

Within this representation, a feasible docking pose corresponds to a set of mutually compatible contacts, which forms a clique in the BIG \cite{kuhl1984combinatorial,Banchi_2020}. The most probable binding configuration is therefore obtained by identifying the clique with the largest total vertex weight, leading to the maximum vertex-weighted clique problem (MVWCP) \cite{Banchi_2020,Ding2024DockingQAOA},

\begin{align}
\label{eq: original MVWCP}
&\max_{x\in{0,1}^N} \sum_i w_i x_i,
\qquad
x_i x_j = 0 \ \text{for incompatible pairs}.
\end{align}

The MVWCP is NP-hard and becomes increasingly challenging as the number of pharmacophore points grows \cite{lucas2014ising,glover2019tutorial}. Nevertheless, this formulation provides a unified framework that couples conformational search and interaction scoring within a single combinatorial optimization problem.

In practical instances, only a subset of pharmacophore points is selected to control graph size while preserving the dominant interaction motifs \cite{Banchi_2020,Ding2024DockingQAOA}. For a given ligand--protein pair, the resulting BIG defines a finite optimization instance whose solution directly specifies a predicted binding pose. This graph-based mapping establishes a bridge between molecular docking and combinatorial optimization, enabling the use of both classical heuristics and quantum algorithms, including Gaussian Boson Sampling and quantum approximate optimization methods \cite{Banchi_2020,yu2023universal,Farhi2014QAOA,Ding2024DockingQAOA,papalitsas2025quantumapproximateoptimizationalgorithms}.

Now, let us consider a BIG which is represented by $\mathcal{G}=(V,E)$, where $E$ represents the set of edges, and $V$ stands for the set of vertices of a graph. Each vertex is assigned with a positive weight $w_j$, which is associated with its pharmacophore potential, and each edge represents an accepted condition between one pair of potential binding sites on the ligand, and the pair of potential binding sites on the peptide \cite{Banchi_2020,Ding2024DockingQAOA}. \tc{Introducing a negative coupling $J_{ij}$ that penalizes every pair of mutually incompatible (non-adjacent) vertices, the $N$-node MVWCP of Eq.~\eqref{eq: original MVWCP} takes exactly the QUBO form of Eq.~\eqref{eq: generic QUBO}, and, through the spin transformation $x_j=\left(z_j+1\right)/2$, the Ising form of Eq.~\eqref{eq: generic Ising} \cite{glover2019tutorial,kochenberger2014unconstrained,lucas2014ising}. The optimal maximum-weighted clique is thereby identified with the ground state of the associated Ising Hamiltonian $\hat{H}=\sum_j\beta_j\hat{h}_j$, which is minimized using the warm-start full-basis encoding scheme of Sec.~\ref{sec:warm-start FBE} \cite{Patti2022MBE,ChaiWarmStart2025,chai2025optimizingquboquantumcomputer}.}

\section{Results}
\label{sec:results}

\subsection{Construction of Two Biologically Relevant Molecular Docking Examples}

To evaluate the graph-based docking formulation with our W-S FBE, we constructed two representative protein–ligand docking instances derived from experimentally resolved Protein Data Bank (PDB) structures: 
\texttt{1stp} and \texttt{9aw2}. In the \texttt{1stp} structure, the bound ligand is 
\emph{biotin}, which forms one of the strongest known non-covalent interactions with 
streptavidin and has been extensively characterized structurally and biophysically 
\cite{weber1989streptavidin,mcconnell2021biotin}. In the \texttt{9aw2} complex, the ligand is \emph{benzamidine}, a classical competitive 
inhibitor of trypsin that binds in the S1 specificity pocket and mimics positively 
charged substrate residues \cite{bode1975trypsin,buch2011trypsin}, such as arginine or lysine. These two systems were selected to provide biologically meaningful examples while exhibiting distinct geometric and interaction characteristics relevant for pharmacophore-based docking.

For each structure, the ligand (biotin for \texttt{1stp} and benzamidine for \texttt{9aw2}) and the corresponding protein binding-site residues were first reduced to pharmacophore representations capturing dominant non-covalent interaction features, including hydrogen-bond donors and acceptors, charged groups, hydrophobic regions, and aromatic centers. Using these pharmacophore points, we constructed labeled distance graphs (LDGs) separately for the ligand and the protein binding site, where vertices correspond to pharmacophore points and edge weights encode pairwise Euclidean distances. This representation preserves the spatial organization of interaction motifs while reducing the dimensionality of the docking problem.

The binding interaction graph (BIG) was then generated by enumerating all candidate contacts between ligand and protein pharmacophore points. Each BIG vertex represents a potential contact pair $(v_l, v_p)$, where $v_l$ belongs to the ligand LDG and $v_p$ belongs to the protein LDG. Consequently, if the ligand and protein contain $n$ and $m$ selected pharmacophore points, respectively, the resulting BIG contains $nm$ vertices.

Edges in the BIG encode compatibility between contacts. Two candidate contacts are connected when they can coexist within a physically feasible docking pose. Compatibility is determined by enforcing geometric consistency between intraligand and intraprotein distances: the distance between two ligand pharmacophores must approximately match the corresponding protein distance within tolerance parameters that account for structural flexibility and interaction-specific uncertainty. This criterion ensures that any clique in the BIG represents a mutually consistent set of contacts.

To incorporate interaction strength, each BIG vertex was assigned a weight determined by the pharmacophore types involved in the contact. These weights reflect relative interaction preferences and bias the optimization toward chemically plausible binding configurations. The resulting graph therefore encodes both geometric feasibility through its edges and interaction energetics through its vertex weights.

Within this representation, a predicted docking pose corresponds to a clique in the BIG, while the most probable pose is obtained by solving the maximum vertex-weighted clique problem. The BIG instances derived from \texttt{1stp} (streptavidin–biotin) and \texttt{9aw2} (trypsin–benzamidine) thus define finite combinatorial optimization problems whose solutions directly specify candidate binding configurations.

The two selected structures provide complementary evaluation regimes. The \texttt{1stp} complex exhibits a well-characterized high-affinity binding mode dominated by hydrogen bonding and hydrophobic interactions typical of the streptavidin–biotin system, and therefore serves as a canonical benchmark for validating graph construction and clique recovery. In contrast, the \texttt{9aw2} system features benzamidine binding in the protease active site with strong electrostatic and hydrogen-bond interactions in the S1 pocket, introducing increased spatial heterogeneity and interaction diversity that lead to different BIG connectivity patterns and optimization landscapes. Together, these examples allow us to assess both correctness and robustness of the docking formulation across biologically relevant scenarios.

\subsection{W-S FBE optimization analysis and hardware solution demonstration on IBM Quantum Processor}

\begin{figure}[h]
    \centering
\includegraphics[width=0.8\columnwidth,draft=false]{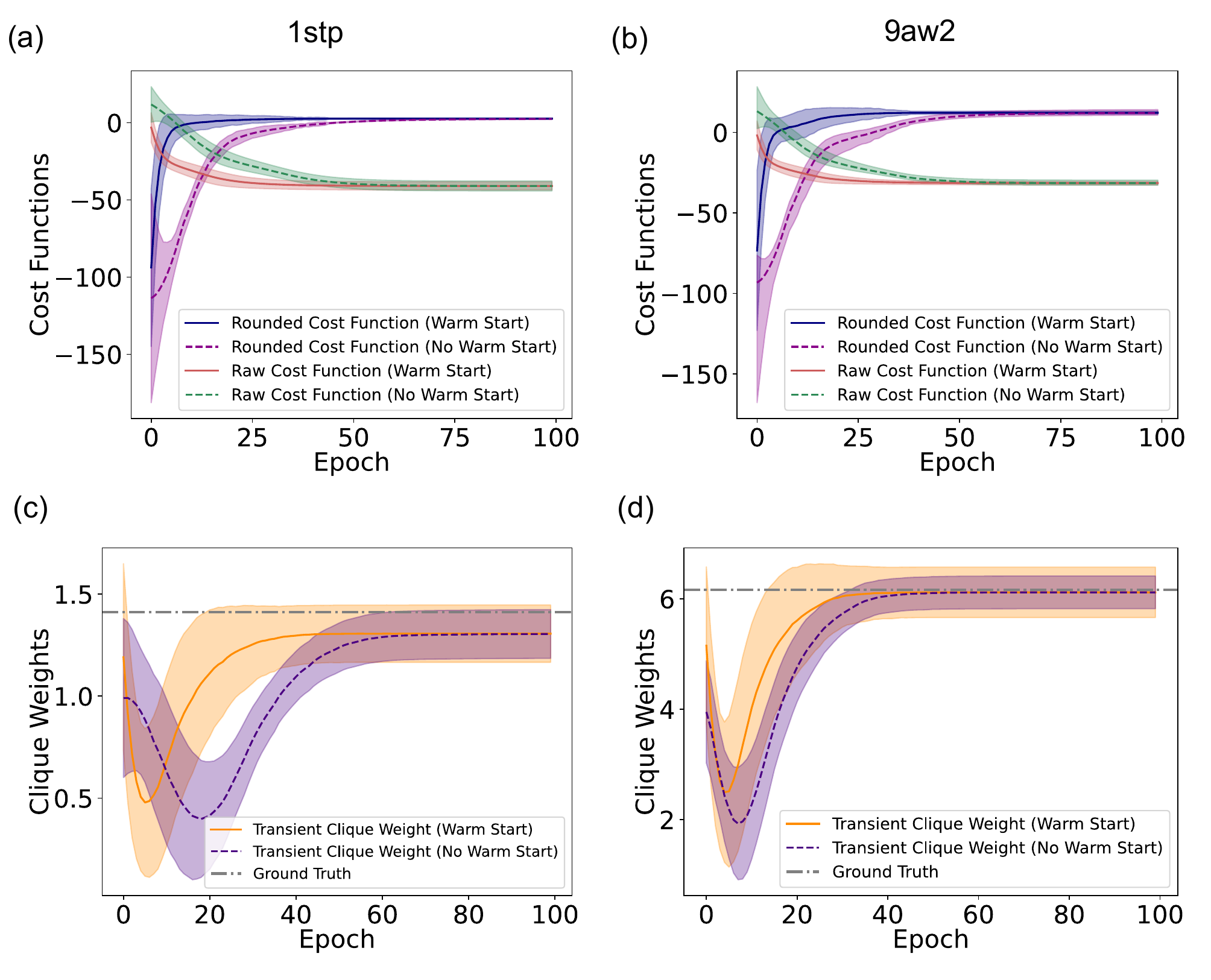}
    \caption{Different loss functions values as a function of training epoch used for checking the convergence of $N=18$ (\texttt{1stp}) in (a) and (c), and $N=14$ (\texttt{9aw2}) in (b) and (d). In panel (a) and (b), the raw cost function (red dashed line) is computed by taking the hyperbolic tangent of each of these observables, i.e., $\tanh{\langle \sigma_{\mu}^i \rangle}$ $(\mu=x,y,z)$ where $i=1,2,\cdots,N$ and insert each for the original variable of the MVWCP cost function, while the rounded cost function (blue solid line) is computed by taking the sign of each $\tanh{\langle \sigma_{\mu}^i \rangle}$ $(\mu=x,y,z)$, and substitute it back to the original MVWCP cost function. In panel (c) and (d), the transient clique weight is calculated for those chosen vertex from the graph at each training epoch (orange solid line), and the grey dot-dashed line represents the ground truth value for each MVWCP graph that is calculated with NetworkX~\cite{hagberg2008exploring}. For all panels, the shaded region for each curve indicates the standard error, and the total number of epoch for training is fixed at $100$. \tc{Other parameters used are: the number of circuit ansatz is fixed to be $3$, the penalty value to be $4$ and the warm-start Trotterized steps are fixed at $15$ for panel (a) and (c), at $10$ for panel (b) and (d).}}
    \label{fig: N=14 and 18 training cost functions}
\end{figure}

We first analyze the optimization behavior of the proposed warm-start full-basis encoding (W-S FBE) on the two docking instances (\texttt{1stp} and \texttt{9aw2}) introduced above. The convergence of the optimization is summarized in Fig.~\ref{fig: N=14 and 18 training cost functions}(a) and (b), where both the raw and the rounded cost functions values are reported as functions of training epoch. In total, as the warm-start is also randomized, we execute $1000$ times of the optimization and compute the average cost function value to eliminate the statistical uncertainty. For the warm-start, we kept the Trotterized steps to be $\mathcal{N}_{\text{T}}=10 (15)$ for the total system size $N=14(18)$ in the case of \texttt{9aw2} (\texttt{1stp}). \tc{The subsequent optimization employs the same type of variational ansatz used throughout this work as described in Fig.~\ref{fig: illustration}(b), where the number of circuit ansatz layers and the penalty value are fixed at 3 and 4, respectively, while the number of warm-start Trotterized steps is set to 15 for \texttt{1stp} and 10 for \texttt{9aw2}.}

For both systems, the raw cost function decreases rapidly during the early training stage, indicating that the variational parameters quickly identify energetically favorable regions, i.e., the ground states of the search space. After this initial descent, the optimization enters a stable regime in which fluctuations remain small enough. On the other hand, the rounded cost function follows the same qualitative trend to the convergence, demonstrating that the relaxed continuous variables translate into the discrete clique assignments. Notably, with warm-start, the convergence to the stable ground states are faster than the case without warm-start for both examples. This behaviour indicates that the warm-start initialization effectively places the variational circuit in a region of parameter space with substantial overlap with low-energy configurations, thereby reducing the number of optimization steps required to reach near-optimal solutions.

\tc{In addition, the transient clique weight provides a complementary view of optimization progress. As shown in Fig.~\ref{fig: N=14 and 18 training cost functions}(c) and (d), \tc{following a small initial decrease from epoch (0) to epoch (1),} the clique weight \tc{then} increases monotonically and approaches the exact optimum (see Fig.~\ref{suppfig: weighted clique} from Appendix.~\ref{sec: MVWCP graphs} for the optimal clique of the total graph for both examples).}
Similar to the rounded and the raw cost function behavior from panel (a) and (b), the warm-start FBE assisted the optimization reaching convergence to optimum at early epoch. 

\begin{figure}[t]
    \centering
    \includegraphics[width=0.7\columnwidth,draft=False]{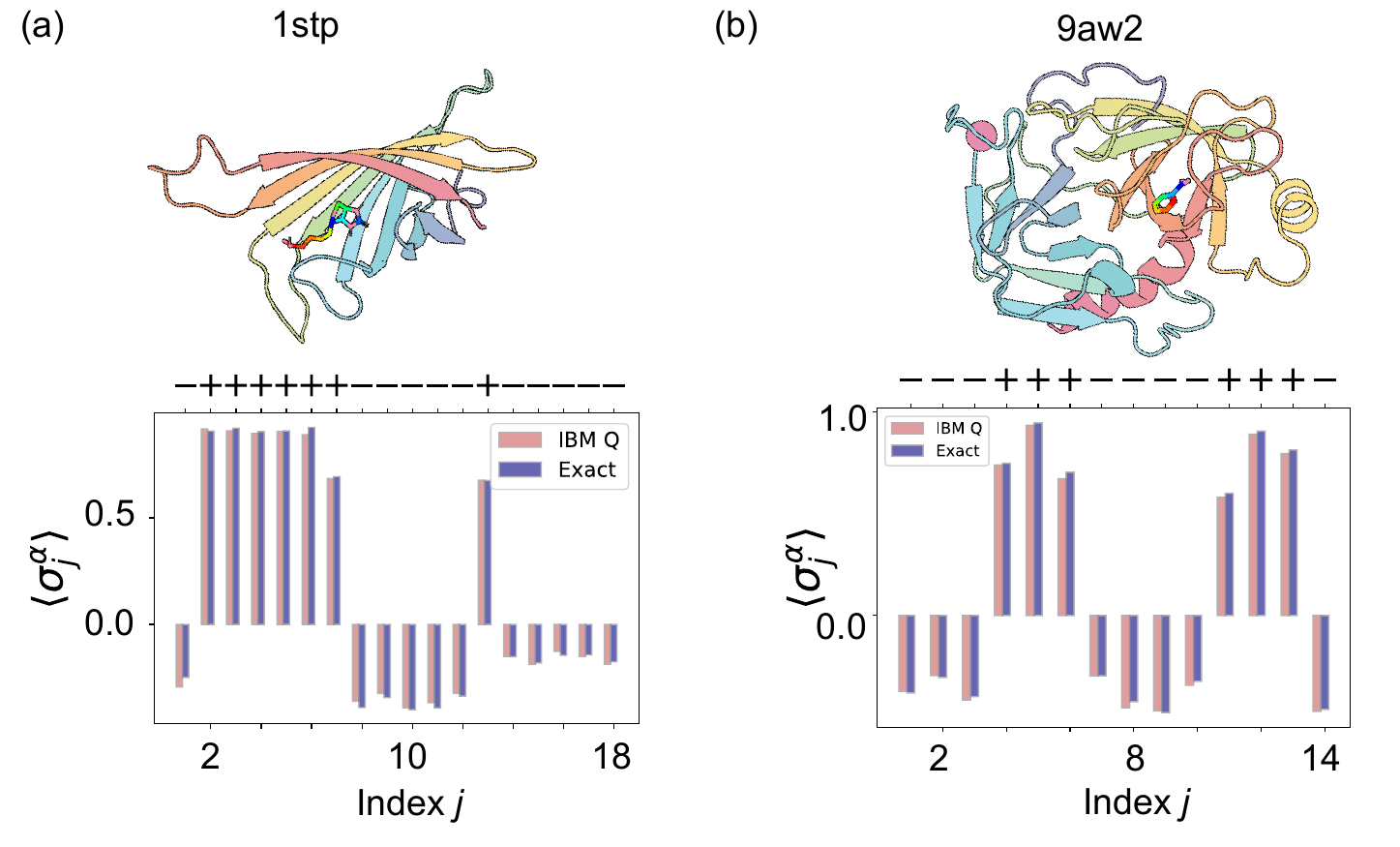}
    \caption{IBM quantum computer demonstration of MVWCP results using full-basis encoding for (a) (PDB ID: \texttt{1stp}) \tc{($N=18$)} and (b) (PDB ID: \texttt{9aw2}). The upper part for each panel is a visualization of the $3$D structure of protein-ligand complex, and the lower part of each panel is the solution of the MVWCP problem obtained from the real IBM hardware. The plus (minus) sign indicate that at the $j$-th vertex of the graph is chosen, and our hardware demonstrations are consistent with both the exact results from classical simulations, as well as the ground truth in Fig.~\ref{suppfig: weighted clique}. The real device results are obtained on a Heron R2 QPU \texttt{ibm\_kingston}. \tc{Other parameters used are: the number of circuit ansatz is fixed to be $3$, and the penalty value is also fixed to be equal to $4$. }For details of the digital quantum simulations, see Appendix.~\ref{sec:details of ibmq}. }
    \label{fig: ibmq results and and docking illustrations with pymol}
\end{figure}

After performing training for both instances using the Adam algorithm~\cite{kingma2017adammethodstochasticoptimization}, the optimized circuits were then executed on IBM quantum hardware \texttt{ibm\_kingston}. The resulting solutions are shown in Fig.~\ref{fig: ibmq results and and docking illustrations with pymol}, and also compared with the exact ideal simulations. For both docking instances, the hardware measurements recover the same vertex selections as classical simulation and match the ground-truth clique structure. For each example, the ligand-protein docking configurations are also visualized using \texttt{PyMOL}~\cite{pymol}. The docking pharmacophore points for \texttt{1stp} and \texttt{9aw2}, and their corresponding ligands are listed in Table.~\ref{table: pharma points} in Appendix. In short, this agreement demonstrates that the W-S FBE ansatz remains robust under realistic hardware gate noise, as well as the readout error, and confirms the feasibility of performing MVWCP-based docking with shallow circuits on current superconducting devices. Overall, these results highlight the warm-start strategy stabilizes training and accelerates convergence, and the full-basis encoding enables resource-efficient and accurate solution recovery on real hardware despite limited circuit depth and number of qubits.

\subsection{Comparison of the Performance between W-S FBE and the ZX-Basis Encoding}

\begin{figure}[h]
    \centering
\includegraphics[width=0.725\columnwidth,draft=False]{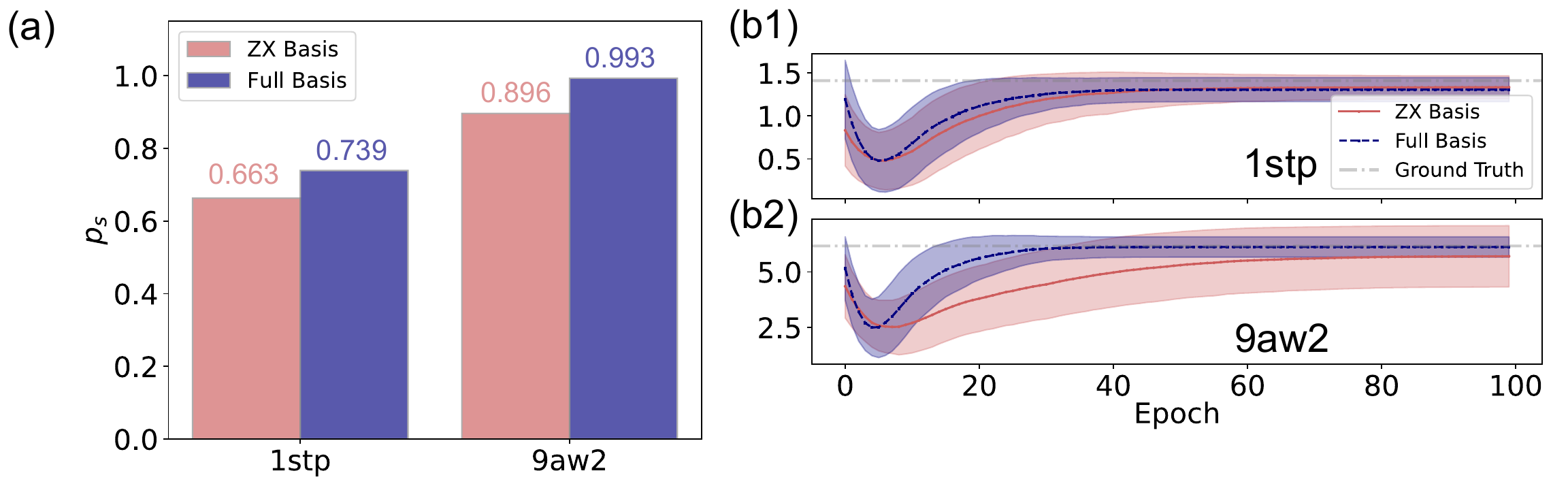}
    \caption{(a) Comparison of successful ratios between ZX basis (red) and full basis encodings (blue) for both \texttt{1stp} ($N=18$) and \texttt{9aw2} ($N=14$). For each approach, it is executed for a total of $1000$ repetitions, and the successful ratio $p_s$ is calculated. The transient clique weight for both approaches are also shown in (b1) for \texttt{1stp} and (b2) for \texttt{9aw2}. In both panels, the horizontal dot-dashed grey line represents the ground truth, and the shaded area stands for the standard error.  \tc{For both cases, the circuit ansatz consists of three layers, while the penalty value is fixed at $4$.}}
    \label{fig: FBE and ZX basis comparison}
\end{figure}

To quantify the benefit of full-basis encoding, we compared W-S FBE against the ZX-basis encoding baseline. Again, the comparison is performed over $1000$ independent repetitions for each docking instance, and the success ratio $p_s$ is evaluated as the probability of obtaining the optimal clique. For ZX-basis encoding, \tc{the same warm-start initialization, variational ansatz structure, optimization protocol, and training budget are employed to ensure a consistent comparison.} Shown in Fig.~\ref{fig: FBE and ZX basis comparison}(a), full-basis encoding \tc{achieves higher success ratios under the optimization settings considered here} for both systems. For the smaller \texttt{9aw2} instance, the success probability approaches unity, while for the larger \texttt{1stp} instance the improvement remains substantial. \tc{We do not interpret this empirical advantage as implying that the theoretical maximum performance of the ZX-basis encoding is necessarily lower than that of FBE. Instead, the ZX-basis encoding uses approximately $50\%$ more qubits for the instances considered here, resulting in a wider quantum circuit and, for the ansatz employed here, a larger variational parameter space. Under the same circuit-depth and training-budget constraints, this increased problem size may make the optimization more difficult and could require deeper circuits or a greater optimization budget to achieve comparable performance. The main novelty and practical advantage of FBE therefore lie in its ability to encode the same optimization problem using fewer qubits by associating multiple classical variables with different Bloch-sphere components. Our results show that this reduced-qubit encoding retains favorable optimization performance under the finite circuit-depth and training budgets considered here.}

The transient clique weight further illustrates and confirms this. Figs.~\ref{fig: FBE and ZX basis comparison}(b1) and (b2) show that FBE reaches the ground-truth clique weight more rapidly and with reduced variance (the shaded regions of the curves) compared with ZX encoding. This indicates that multi-axis observables provide a smoother optimization landscape and improve stability during training. Additional insight is provided by the comparison of raw and rounded objectives (see Fig.~\ref{suppfig:other_functions_ZX_and_full_basis} in Appendix.~\ref{sec: additional FBE and ZX comparison}).

Taken together, these results demonstrate that exploiting all three Pauli bases improves optimization reliability, increases success probability, and enhances convergence stability, particularly for moderately sized docking graphs where qubit resources are constrained.

\subsection{Effect of Penalty Values and the Circuit Ansatz}

\begin{figure}[h]
    \centering
    \includegraphics[width=0.8\columnwidth,draft=False]{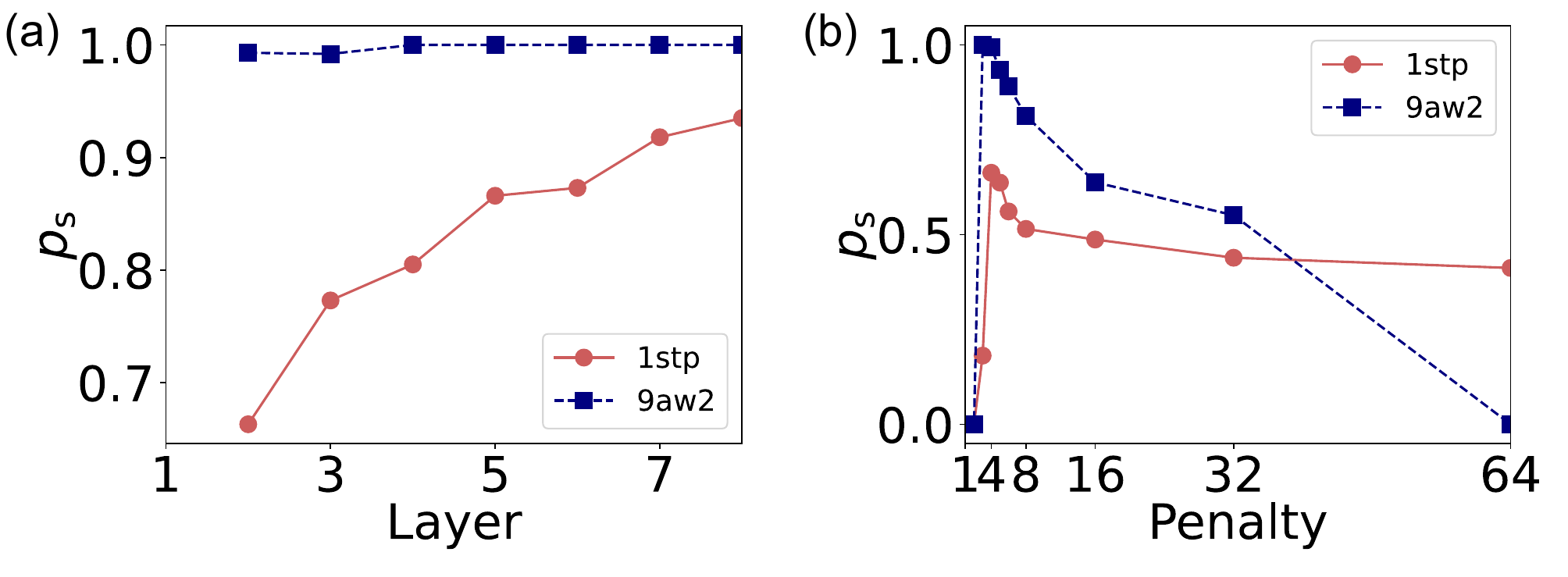}
    \caption{Effect of number of (a) the variational quantum circuit layers and (b) the penalty value for \texttt{1stp} ($N=18$, red) and \texttt{9aw2, red} ($N=14$) on the successful ratio $p_s$. For both panels, it is executed for a total of $1000$ repetitions, and the successful ratio $p_s$ is calculated. }
    \label{fig: effect of parameters}
\end{figure}

Finally, we investigate how hyperparameters can influence the performance, focusing on circuit depth and penalty strength \tc{for the W-S FBE approach}. The success ratio $p_s$, defined as the successful optimization converging to the optimum, as a function of the number of variational layers is shown in Fig.~\ref{fig: effect of parameters}(a). Increasing circuit depth improves performance for $N=18$ (\texttt{1stp}), consistent with the expectation that additional layers enhance expressibility. On the other hand, however, the improvement saturates beyond a modest number of layers for $N=14$ (\texttt{9aw2}), indicating that shallow circuits already capture the dominant structure of the optimization landscape for relatively smaller system sizes \tc{where the number of nodes on the graph for the MVWCP optimization is small}.

\tc{The influence of the penalty coefficient is depicted in Fig.~\ref{fig: effect of parameters}(b).} The penalty term enforces incompatibility constraints in the MVWCP formulation and therefore balances feasibility against the (raw) cost function minimization. We observe that for both cases, excessively small penalties lead to constraint violations and reduced success probability, while overly large penalties suppress exploration and degrade optimization. In particular, for smaller sizes (\texttt{9aw2}), larger penalty severely affect the success ratio compared with larger sizes due to less possible choices of bitstrings. In principle, an intermediate regime yields the best performance, \tc{suggesting that penalty tuning is essential for stable training to obtain the optimal solution of MVWCP.}. 

Importantly, the optimal hyperparameter region is consistent across both docking instances (Fig.~\ref{fig: effect of parameters}), indicating that the method does not require extensive problem-specific hyperparameter calibration and tuning. This robustness is particularly relevant for practical docking pipelines, where many graph instances must be solved sequentially for biologically-relevant practical reasons. Overall, it is shown that W-S FBE maintains strong performance across a broad range of circuit depths and penalty values, while retaining the shallow-circuit advantage required for NISQ hardware. The observed trade-offs highlight the importance of balancing expressibility, constraint enforcement, and noise sensitivity when designing variational quantum optimization algorithms for molecular docking.

\section{Conclusions and Discussions}
\label{sec:conclusions}
\tc{
We introduced a resource-efficient hybrid quantum--classical framework for 
molecular docking by mapping pharmacophore compatibility onto a maximum 
vertex-weighted clique problem and solving the resulting combinatorial 
optimization problem using warm-start full-basis encoding (W-S FBE). 
The central technical contribution is a new encoding resource trade-off: 
by associating classical variables with the three Bloch-vector components 
of each qubit, FBE can encode up to three variables per qubit and reduce an 
$N$-variable problem to only $\lceil N/3\rceil$ qubits. The variables hosted 
on the same qubit are coupled through the Bloch-sphere constraint, producing 
a structured continuous relaxation of the original discrete optimization 
problem. Importantly, we prove that at least one global minimizer of the FBE 
objective can always be represented by a pure product state. This result 
establishes that entanglement is not required to represent an optimal solution 
and provides a rigorous foundation for optimizing the FBE objective using a 
unitary variational circuit. Combined with a stochastic 
imaginary-time-evolution-inspired warm start and a shallow variational ansatz, 
W-S FBE achieves stable convergence for the biologically relevant docking 
instances considered here. Under matched ansatz structures and finite 
optimization budgets, FBE also exhibits higher empirical success probabilities 
than the ZX-basis encoding while requiring fewer qubits. We emphasize that this 
observation reflects a finite-resource advantage under the settings studied 
here, rather than a fundamental limitation on the theoretical performance of 
the ZX-basis encoding. Finally, execution on an IBM superconducting quantum 
processor recovers the same optimal clique structures as the noiseless 
simulations, demonstrating the feasibility of the proposed compressed encoding 
on current quantum hardware.
}

\tc{
The significance of the present approach lies not in replacing established classical docking pipelines, but in providing a compact quantum-assisted solver for their discrete contact-selection subproblems. After the ligand and protein binding site have been reduced to pharmacophore representations, the selection of a mutually compatible set of interactions becomes a constrained combinatorial problem. Conventional basis encodings typically assign a separate qubit to each binary variable, whereas FBE exploits the full Bloch vector to trade qubit count for a geometrically constrained relaxation. This distinction is particularly relevant for near-term devices, for which the number of available high-fidelity qubits and the trainability of increasingly large variational circuits remain major bottlenecks. The present results therefore establish FBE as a resource-efficient and experimentally implementable encoding strategy for quantum-assisted molecular docking, rather than as evidence of \textit{quantum advantage} over optimized classical methods.
}

Several challenges remain. The present study considers moderate graph sizes for which exact classical verification is still possible, and the results should therefore be interpreted as evidence of feasibility rather than quantum advantage. Scaling to more realistic docking problems will require more chemically accurate and computationally efficient construction of binding interaction graphs, improved variational ansatz design, adaptive penalty and hyperparameter selection, noise-aware optimization, and tighter integration between classical preprocessing and quantum optimization. Future work will extend the framework to larger and more heterogeneous interaction graphs, flexible docking scenarios, and alternative interaction-scoring models. It will also explore tensor-network-inspired sampling methods ~\cite{Liu2018Born,benedetti2019generative,Du2020Expressive, rudolph2024trainability,harada2025tensor, bendov2025regularizedsecondorderoptimizationtensornetwork}, stack-operations-based acceleration~\cite{Zhang2022}, surrogate-assisted optimization~\cite{surrogate2025}, and hardware-aware circuit design ~\cite{Wecker2015progress,Kandala2017hardware,Cincio2018adaptive, Zulehner2018efficient,Moll2018quantum,Li2020trotterization, Sharma2020error,pozzi2020usingreinforcementlearningperform, Li2021hardware,Bravyi2021circuit}. It is also worth exploring the non-variational quantum algorithms to tackle this problem. The decisive scaling test will be whether the qubit compression introduced by FBE continues to improve solution quality per physical qubit as docking graphs become larger, denser, and more chemically realistic.

\section*{Acknowledgements}
We thank Hwee Kuan Lee, Haiyue Kang, Thiparat Chotibut, Pranav Kalidindi, Derek Kim, Shashvat Shukla, V Vijendran, Syed Assad, Lorcan Conlon and Ping Koy Lam for fruitful discussions. \tc{This work is supported by the National Research Foundation, Singapore, through the National Quantum Office, hosted in A*STAR, under the Quantum Engineering Programme 3.0 Funding Initiative (W24Q3D0002), the Hybrid Quantum-Classical Computing (HQCC) 1.0 Funding Initiative (S24Q7D7001), and the Advanced Quantum Algorithms and Solutions (AQAS) Funding Initiative (S25Q9DA001).} T.~C. also acknowledges partial support from A*STAR Q.~Inc Strategic Research and Translational Thrust (SRTT). T.~C. acknowledges the use of IBM Quantum Credits for this work when he was at National University of Singapore. The tensor network states calculation in this work is performed using Tensorly Quantum~\cite{Patti2022MBE}. The network graph in this work is generated using NetworkX~\cite{hagberg2008exploring}. The computational work for this article was partially performed on resources of the National Supercomputing Centre, Singapore (\url{https://www.nscc.sg/}), and was partially supported by the A*STAR Computational Resource Centre through the use of its high performance computing facilities.

\bibliography{ref}

\begin{thebibliography}{82}
\bibitem [{rou()}]{roundupnote}%
  \BibitemOpen
  \href@noop {} {}\bibinfo {note} {Here, the symbol $\lceil \cdots \rceil$
  represents the smallest integer greater than or equal to the input. As a
  result,if $M$ is not divisible by $3$, then for the last qubit, some
  observables are not utilized to substitute the classical variable from the
  combinatorial problem.}\BibitemShut {Stop}%
\end{thebibliography}
\newpage
\appendix

\section{\tc{Proof of the ground state of the Hamiltonian as a product state}}
\label{sec:proofofgroundstate}

\tc{In this section, we prove the statement made in the main text that the ground state of $\mathcal{L}_{\mathrm{FBE}}^{\mathrm{raw}}$ is essentially a product state. This result provides a rigorous foundation for the use of a unitary quantum circuit as the optimization ansatz in this work.
}
\tc{
We first define some notation and assumptions. Define}
\tc{
\begin{equation}
    s_q^\alpha(\bm{\vartheta})
    := \langle\psi(\bm{\vartheta})\rvert\sigma_q^\alpha
       \lvert\psi(\bm{\vartheta})\rangle,
    \qquad
    \alpha\in\{x,y,z\},\quad q\in\{1,\ldots,m\}.
\end{equation}
}
\tc{
For each vertex index $v$ from the graph, let
\begin{equation}
    x_v(\bm{\vartheta})
    = \tanh\!\left(s_{q(v)}^{\alpha(v)}(\bm{\vartheta})\right)
    \equiv \Phi\!\left(s_{q(v)}^{\alpha(v)}\right),
    \qquad \Phi(t):=\tanh t.
\end{equation}
}
\tc{
The raw FBE loss is
\begin{equation}
    \mathcal{L}_{\mathrm{FBE}}^{\mathrm{raw}}
    = -\sum_{u\neq v}J_{uv}x_u x_v-\sum_u w_u x_u.
\end{equation}
}
\tc{
For each qubit $q$, define its local Bloch vector by
\begin{equation}
    \bm{s}_q=(s_q^x,s_q^y,s_q^z),
    \qquad \lVert\bm{s}_q\rVert\leq 1.
\end{equation}
}
\tc{
Assume that the embedding
\begin{equation}
    \Pi(v)=\bigl(q(v),\alpha(v)\bigr)
\end{equation}
is injective, so that each classical variable is mapped to a unique
qubit-axis pair as above.
}

\medskip
\begin{tcolorbox}
\tc{
\noindent\textbf{Theorem.}
Minimizing $\mathcal{L}_{\mathrm{FBE}}^{\mathrm{raw}}$ over arbitrary $m$-qubit states is equivalent to
minimizing it over pure product states. More precisely,
\begin{equation}
    \min_{\mathrm{arbitrary}\,\rho}\mathcal{L}_{\mathrm{FBE}}^{\mathrm{raw}}(\rho)
    =\min_{\mathrm{ product}\,\rho}\mathcal{L}_{\mathrm{FBE}}^{\mathrm{raw}}(\rho)
    =\min_{\mathrm{pure\ state}}\mathcal{L}_{\mathrm{FBE}}^{\mathrm{raw}}(\rho)
\label{eq:three-minima}
\end{equation}
Thus, at least one global minimizer of the raw FBE loss is a pure
product state.}
\end{tcolorbox}
\medskip

\tc{
\noindent\textit{Proof.}
We first establish the first equality in Eq.~\eqref{eq:three-minima}.
Given an arbitrary $m$-qubit density matrix $\rho$, define
\begin{equation}
    s_q^\alpha=\operatorname{Tr}\!\left(\rho\,\sigma_q^\alpha\right).
\end{equation}
The reduced density matrix of qubit $q$ is then
\begin{equation}
    \rho_q
    =\operatorname{Tr}_{\bar q}(\rho)
    =\frac{1}{2}\left(I+\bm{s}_q\cdot\bm{\sigma}_q\right),
    \qquad
    \bm{\sigma}_q=(\sigma_q^x,\sigma_q^y,\sigma_q^z).
\end{equation}
\textcolor{black}{Here, the reduced density matrix for each individual qubit $q$ is written as a general form which can be either a pure state of a mixed state. }The loss depends on $\rho$ only through the quantities
\begin{equation}
    x_v(\rho)=\tanh\!\left(s_{q(v)}^{\alpha(v)}\right).
\end{equation}
Now we construct the product density matrix $\widetilde\rho$
\begin{equation}
    \widetilde\rho=\bigotimes_{q=1}^{m}\rho_q.
\end{equation}
\textcolor{black}{Without loss of generality, $\widetilde\rho$, which we term as `product $\rho$' in Eq.~\eqref{eq:three-minima}, can also be either a pure state of a mixed state. }It has exactly the same single-qubit expectation values as $\rho$:
\begin{equation}
    \operatorname{Tr}\!\left(\widetilde\rho\,\sigma_q^\alpha\right)
    =s_q^\alpha
    =\operatorname{Tr}\!\left(\rho\,\sigma_q^\alpha\right).
\end{equation}
Consequently,
\begin{equation}
    \mathcal{L}_{\mathrm{FBE}}^{\mathrm{raw}}(\rho)=\mathcal{L}_{\mathrm{FBE}}^{\mathrm{raw}}(\widetilde\rho),
\end{equation}
which proves the first equality.}
\tc{
For the second equality, let
\begin{equation}
    \mathbb{B}^3:=\{\bm{s}\in\mathbb{R}^3:\lVert\bm{s}\rVert\leq 1\},
    \qquad
    \mathbb{S}^2:=\{\bm{s}\in\mathbb{R}^3:\lVert\bm{s}\rVert=1\}.
\end{equation}
Every product state of the above form can be written as
\begin{equation}
    \widetilde\rho(\bm{s})
    =\bigotimes_{q=1}^{m}
      \frac{1}{2}\left(I+\bm{s}_q\cdot\bm{\sigma}_q\right),
    \qquad
    \bm{s}=(\bm{s}_1,\ldots,\bm{s}_m)\in(\mathbb{B}^3)^m.
\end{equation}
Hence
\begin{equation}
    \min_{\widetilde\rho}\mathcal{L}_{\mathrm{FBE}}^{\mathrm{raw}}(\widetilde\rho)
    =\min_{\bm{s}\in(\mathbb{B}^3)^m}
       \mathcal{L}_{\mathrm{FBE}}^{\mathrm{raw}}\!\left(\widetilde\rho(\bm{s})\right).
\label{eq:bloch-minimization}
\end{equation}
Because $(\mathbb{B}^3)^m$ is compact and $\mathcal{L}_{\mathrm{FBE}}^{\mathrm{raw}}$ is continuous, a global
minimum exists based on the extreme value theorem.}

\tc{
We next examine the conditional dependence on the one-qubit case. Fix all
Bloch vectors except $\bm{s}_q=(s_q^x,s_q^y,s_q^z)$.
Since $\mathcal{L}_{\mathrm{FBE}}^{\mathrm{raw}}$ is quadratic in the variables $x_v$, and each qubit axis
hosts at most one variable, its dependence on $\bm{s}_q$ can be written
as
\begin{equation}
    g(\bm{s}_q)
    =c+\sum_{\alpha}a_\alpha\Phi(s_q^\alpha)
     +\sum_{\alpha<\beta}
       b_{\alpha\beta}\Phi(s_q^\alpha)\Phi(s_q^\beta),
\label{eq:conditional-g}
\end{equation}
where $\alpha,\beta\in\{x,y,z\}$. Here $a_\alpha$ contains the linear
weight assigned to the variable associated with $(q,\alpha)$, together
with its interactions with variables on other qubits, while
$b_{\alpha\beta}$ describes interactions between variables hosted on
different axes of the same qubit. Here, we remark that there is no second-order term of
the form $\Phi(s_q^\alpha)^2$.}

\medskip

\begin{tcolorbox}
\tc{
\noindent\textbf{Lemma.}
A function of the form in Eq.~\eqref{eq:conditional-g}, when minimized
over $\mathbb{B}^3$, has a global minimizer on the boundary $\mathbb{S}^2$.}
\end{tcolorbox}

\medskip

\tc{
\noindent\textit{Proof.}
This is done through proof by contradiction. If $g$ is constant, the statement is immediate and trivial. Suppose instead that
$g$ is not a constant and, toward a contradiction, that it has a local minimum $\bm{s}^*$ with $\lVert\bm{s}^*\rVert<1$. Below, we suppress the qubit index for brevity, and extend the coefficients symmetrically by
$b_{\beta\alpha}=b_{\alpha\beta}$. Then
\begin{equation}
\label{eq:stationarycondition}
    \frac{\partial g\left(\bm{s}\right)}{\partial s_\alpha}
    =\Phi'(s_\alpha)
      \left(a_\alpha+\sum_{\beta\neq\alpha}
      b_{\alpha\beta}\Phi(s_\beta)\right)
    \equiv \Phi'(s_\alpha)h_\alpha(\bm{s}).
\end{equation}
Since
\begin{equation}
    \Phi(s_\alpha)=\tanh(s_\alpha),
    \qquad
    \Phi'(s_\alpha)=\operatorname{sech}^2(s_\alpha)>0,
\end{equation}
the condition from the minimization from Eq.~\eqref{eq:stationarycondition} implies
\begin{equation}
    h_\alpha(\bm{s}^*)=0
    \qquad\forall\,\alpha.
\end{equation}
The diagonal entries of the Hessian of the partial derivative are
\begin{equation}
    H_{\alpha\alpha}
    =\frac{\partial^2g}{\partial s_\alpha^2}
    =\Phi''(s_\alpha)h_\alpha(\bm{s}),
\end{equation}
so $H_{\alpha\alpha}(\bm{s}^*)=0$. Its off-diagonal entries are
\begin{equation}
    H_{\alpha\beta}
    =b_{\alpha\beta}\Phi'(s_\alpha)\Phi'(s_\beta),
    \qquad \alpha\neq\beta.
\end{equation}
At a local minimum, $H(\bm{s}^*)$ is positive semi-definite. Every
$2\times2$ principal minor must therefore satisfy
\begin{equation}
    0\leq
    \det\begin{pmatrix}
        0&H_{\alpha\beta}\\
        H_{\alpha\beta}&0
    \end{pmatrix}
    =-H_{\alpha\beta}^2,
\end{equation}
which forces $H_{\alpha\beta}=0$ for all $\alpha\neq\beta$. Since
$\Phi'>0$, it follows that $b_{\alpha\beta}=0$ for all
$\alpha\neq\beta$. The equations $h_\alpha(\bm{s}^*)=0$ then imply
$a_\alpha=0$ for all $\alpha$, so $g(\bm{s})=c$ is constant for any $\alpha$. Therefore, this is a contradiction to the assumption that $g(\bm{s})$ is not a constant in the beginning.
}

\tc{
Thus a non-constant $g$ has no local minimum. Since $\mathbb{B}^3$ is
compact, a global minimum exists and can be obtained on its boundary
$\mathbb{S}^2$ according to the extreme value theorem.
}
\hfill$\square$\par
\medskip

\tc{
We now apply the lemma to each qubit. Let
\begin{equation}
    \bm{s}^*=(\bm{s}_1^*,\ldots,\bm{s}_m^*)
\end{equation}
be a global minimizer over $(\mathbb{B}^3)^m$. Holding
$\bm{s}_2^*,\ldots,\bm{s}_m^*$ fixed, the vector $\bm{s}_1^*$ minimizes
the corresponding conditional function. By the lemma, it may be
replaced by a minimizer on $\mathbb{S}^2$ without changing the global minimum
value. Repeating this argument from qubits $2$ through qubit $m$ gives a
global minimizer satisfying
\begin{equation}
    \lVert\bm{s}_i^*\rVert=1,
    \qquad i=1,\ldots,m.
\end{equation}
Therefore Eq.~\eqref{eq:bloch-minimization} can be written as
\begin{equation}
    \min_{\bm{s}\in(\mathbb{B}^3)^m}
       \mathcal{L}_{\mathrm{FBE}}^{\mathrm{raw}}\!\left(\widetilde\rho(\bm{s})\right)
    =
    \min_{\bm{s}\in(\mathbb{S}^2)^m}
       \mathcal{L}_{\mathrm{FBE}}^{\mathrm{raw}}\!\left(\widetilde\rho(\bm{s})\right).
\end{equation}
}
\tc{
Finally, for every $q$ with $\lVert\bm{s}_q^*\rVert=1$, choose a pure
state $\lvert\psi_q\rangle$ satisfying
\begin{equation}
    \lvert\psi_q\rangle\!\langle\psi_q\rvert
    =\frac{1}{2}\left(I+\bm{s}_q^*\cdot\bm{\sigma}_q\right).
\end{equation}
Then
\begin{equation}
    \lvert\Psi_{\mathrm{prod}}\rangle
    =\bigotimes_{q=1}^{m}\lvert\psi_q\rangle
\end{equation}
is a pure product state attaining the global minimum. This completes
the proof.}
\hfill$\square$\par

\section{Details of Digital Quantum Simulations on an IBM Quantum Processor}
\label{sec:details of ibmq}

\begin{figure}[t]
    \centering
\includegraphics[width=0.75\columnwidth,draft=False]{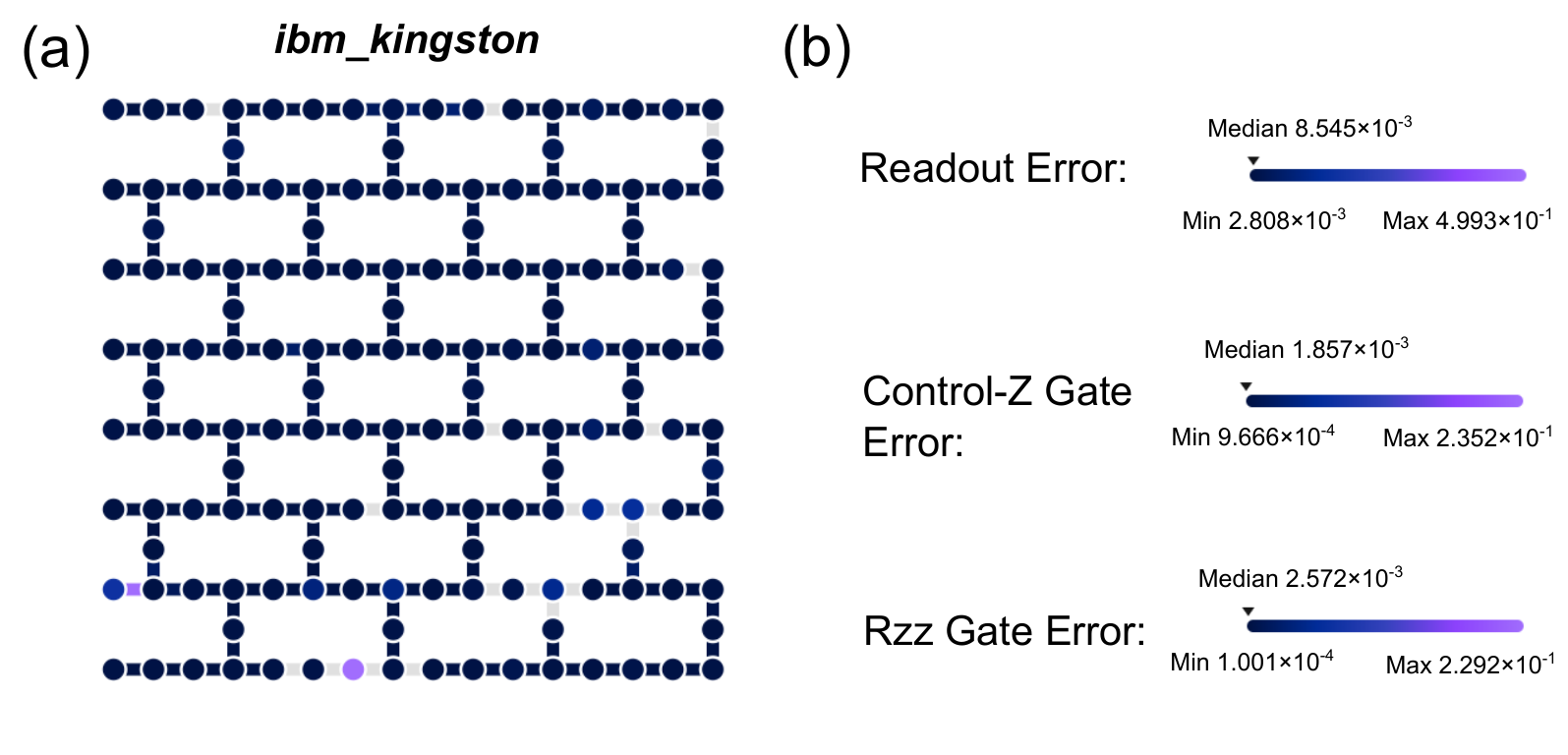}
    \caption{Details of the IBM quantum processor: (a) Schematic of the \texttt{ibm\_kingston} processor, belonging to the \tc{$156$-qubit} Heron r$2$ family~\cite{ibm_quantum_2025}. Each solid circle denotes an individual qubit, with bonds indicating physical couplings between them. The color shading of both circles and bonds reflects the magnitude of gate errors (shown in panel (b) as well): lighter tones correspond to higher error rates. We select the qubits and their connecting edge selected for the algorithm demonstrated in the main text by leveraging the calibration data of the device~\cite{ibm_quantum_2025}. (b) Control-Z as well as RZZ gate infidelities, together with the readout assignment (measurement) average, maximum and minimum error for the \texttt{ibm\_kingston} processor. }
    \label{suppfig: details of ibmq}
\end{figure}

For all the demonstration results presented in the main text, we executed the quantum circuits on the IBM quantum processor~\cite{ibm_quantum_2025}, which is a state-of-the-art quantum computing platform that leverages superconducting qubits. All the quantum circuits are constructed via Qiskit API, and executed via Qiskit Runtime service~\cite{Qiskit2024}.

All physical demonstration of results on quantum circuits reported in the main text were executed and obtained on IBM superconducting quantum processors~\cite{ibm_quantum_2025} using the Qiskit Runtime service~\cite{Qiskit2024}. Circuits were transpiled with hardware-aware optimization to respect connectivity constraints and minimize two-qubit gate overhead. In this work, all hardware simulations are performed as a one-dimensional layout of quantum circuit. Fig.~\ref{suppfig: details of ibmq} summarizes the device characteristics relevant to this study. We employed a Heron-family processor (ibm\_kingston) and selected qubit subsets based on calibration metrics, including readout error, controlled-$Z$ infidelity and the $RZZ$ gate error. Qubit allocation was chosen to minimize cumulative two-qubit error along the interaction topology required by the variational ansatz. This hardware-aware selection ensures that algorithmic performance reflects intrinsic encoding properties rather than suboptimal qubit placement. In addition to this, the expectation values of Pauli observables [$\langle \sigma_\alpha \rangle (\alpha=x,y,z)$] were obtained using the Estimator primitive, enabling direct evaluation of single-qubit observables required by the full-basis encoding. See Table.~\ref{table: measured Pauli strings} for the measured Pauli strings and qubit arrangements for both full basis and ZX basis of \texttt{1stp} ($N=18$) and \texttt{9aw2} ($N=14$). Shot allocation was kept uniform across measurement bases to avoid bias in the multi-axis relaxation. Error mitigation was limited to standard readout mitigation so as to preserve a realistic assessment of algorithm robustness.

\begin{figure}[h]
    \centering
    \includegraphics[width=0.75\columnwidth,draft=False]{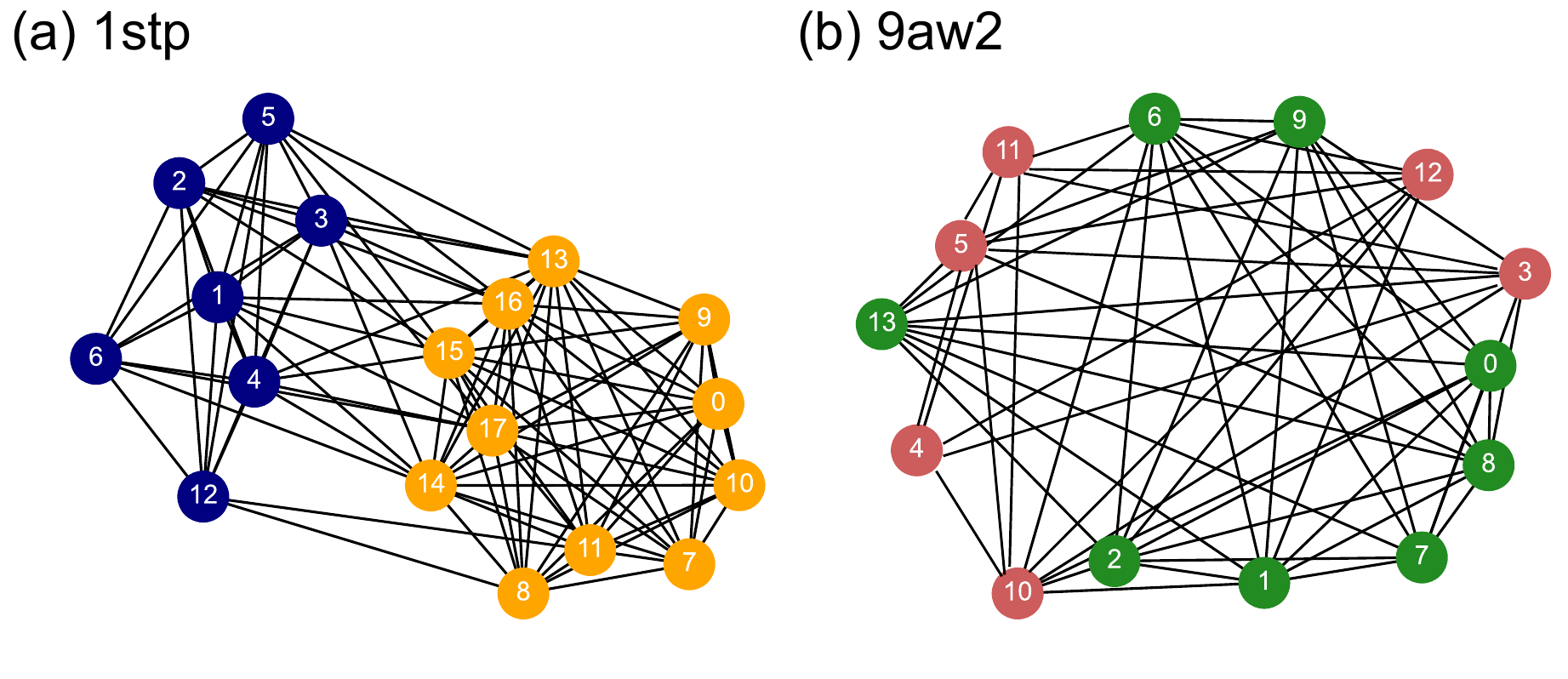}
    \caption{The maximum vertex-weighted clique problem (MVWCP) for (a) \texttt{1stp} and (b) \texttt{9aw2}. The labeling for the vertex starts from $0$. The orange (green) color in panel (a) (panel (b)) indicates the solution for the MVWCP within each graph.}
    \label{suppfig: weighted clique}
\end{figure}

\begin{table*}[h]

  \centering

  \begin{minipage}{1.0\linewidth}
    \centering
    {\bfseries\strut \texttt{1stp} ($N=18$)}

    \begin{tabular}{ c c c }
      \toprule
      \bfseries Ligand Atom & \bfseries Residue Name & \bfseries Atom Name \\
      \midrule
       C3   & TRP79 & HN \\
       C3   & TRP79 & CG \\
       C3   & TRP79 & CA \\
       C3   & TRP79 & N \\
       C5   & TRP79 & N \\
       C8   & VAL47 & O \\
       C5   & TRP79 & CG \\
       C5   & TRP79 & CA \\
       C3   & TRP79 & C \\
       C5   & TRP79 & C \\
       C5   & TRP79 & HN \\
      \bottomrule
    \end{tabular}

  \end{minipage}%
  \quad 
  \begin{minipage}{\linewidth}
    \centering
    {\bfseries\strut \texttt{9aw2} ($N=14$)}

    \begin{tabular}{ c c c }
      \toprule
      \bfseries Ligand Atom & \bfseries Residue Name & \bfseries Atom Name \\
      \midrule
      C1 & GLN174 & N \\
      C1 & GLN174 & SG \\
      C1 & VAL191 & CG1 \\
      C1 & CYS 173 & CB \\
      N1 & GLN174 & SG \\
      N1 & GLN174 & N\\
      N1 & VAL191 & CG1\\
      N1 & CYS 173 & CB \\
      \bottomrule
    \end{tabular}
  \end{minipage}
  \caption{Lists of molecular docking pharmacophore points for \texttt{1stp} and \texttt{9aw2}. In the column representing the ligand atom, the number refers to the intrinsic atom index within the ligand. In the column for the residue name, the number refers to the index for the amino acid residue within the peptide chain sequence. Other notations include TRP: Tryptophan, VAL: Valine, GLN: Glutamic Acid, CYC: Cysteine, HN: backbone amide hydrogen, CG: carbon connecting to indole ring, CA: alpha carbon, N: backbone amide nitrogen, O: carbonyl oxygen, C: carbonyl carbon, SG: sulfur atom (thiol group), CG1: first gamma carbon (methyl group \#1), CB: beta carbon (first carbon in side chain, bonded to CA). }\label{table: pharma points}
\end{table*}

\begin{table*}[h]
  \centering

  \begin{minipage}{1.0\linewidth}
    \centering
    {\bfseries\strut \texttt{1stp} ($N=18$)}

    \begin{tabular}{ c c c}
      \toprule
      \bfseries Full Basis &\bfseries ZX Basis  & \bfseries Variable [$x_i$] \\
      \midrule
      \itshape $IIIIIZ$ &\itshape $IIIIIIIIZ$ & {$x_1$}\\
      \itshape $IIIIZI$ &\itshape $IIIIIIIZI$ & {$x_2$}\\
      \itshape $IIIZII$ &\itshape $IIIIIIZII$ & {$x_3$}\\
      \itshape $IIZIII$ &\itshape $IIIIIZIII$ & {$x_4$}\\
      \itshape $IZIIII $ &\itshape $IIIIZIIII$ & {$x_5$}\\
      \itshape $ZIIIII $ &\itshape $IIIZIIIII$ & {$x_6$}\\
      \itshape $IIIIIX $ &\itshape $IIZIIIIII$ & {$x_7$}\\
      \itshape $IIIIXI $ &\itshape $IZIIIIIII$ & {$x_8$}\\
      \itshape $IIIXII $ &\itshape $ZIIIIIIII$ & {$x_9$}\\
      \itshape $IIXIII $ &\itshape $IIIIIIIIX$ & {$x_{10}$}\\
      \itshape $IXIIII $ &\itshape $IIIIIIIXI$ & {$x_{11}$}\\
      \itshape $ XIIIII $ &\itshape $IIIIIIXII$  & {$x_{12}$}\\
      \itshape $IIIIIY $ &\itshape $IIIIIXIII$ & {$x_{13}$}\\
      \itshape $ IIIIYI $ &\itshape $IIIIXIIII$ & {$x_{14}$}\\
      \itshape $ IIIYII $ &\itshape $IIIXIIIII$ & {$x_{15}$}\\
      \itshape $ IIYIII $ &\itshape $IIXIIIIII$ & {$x_{16}$}\\
      \itshape $ IYIIII $ &\itshape $IXIIIIIII$ & {$x_{17}$}\\
      \itshape $ YIIIII $ &\itshape $XIIIIIIII$ & {$x_{18}$}\\
      \bottomrule
    \end{tabular}

  \end{minipage}%
  \quad 
  \begin{minipage}{1.0\linewidth}
    \centering
    {\bfseries\strut \texttt{9aw2} ($N=14$)}
    
    \begin{tabular}{c c c}
      \toprule
      \bfseries Pauli String &\bfseries ZX Basis &\bfseries Variable [$x_i$]\\
      \midrule
      \itshape $IIIIZ$ &\itshape $IIIIIIZ$ & {$x_1$}\\
      \itshape $IIIZI$ &\itshape $IIIIIZI$ & {$x_2$}\\
      \itshape $IIZII$ &\itshape $IIIIZII$ & {$x_3$}\\
      \itshape $IZIII$ &\itshape $IIIZIII$ & {$x_4$}\\
      \itshape $ZIIII$ &\itshape $IIZIIII$ & {$x_5$}\\
      \itshape $IIIIX$ &\itshape $IZIIIII$ & {$x_6$}\\
      \itshape $IIIXI$ &\itshape $ZIIIIII$ & {$x_7$}\\
      \itshape $IIXII$ &\itshape $IIIIIIX$ & {$x_8$}\\
      \itshape $IXIII$ &\itshape $IIIIIXI$ & {$x_{9}$}\\
      \itshape $XIIII$ &\itshape $IIIIXII$ & {$x_{10}$}\\
      \itshape $IIIIY$ &\itshape $IIIXIII$ & {$x_{11}$}\\
      \itshape $IIIYI$ &\itshape $IIXIIII$ & {$x_{12}$}\\
      \itshape $IIYII$ &\itshape $IXIIIII$ & {$x_{13}$}\\
      \itshape $IYIII$ &\itshape $XIIIIII$ & {$x_{14}$}\\
      \bottomrule
    \end{tabular}
  \end{minipage}%
  \caption{Tables of the measured Pauli strings for both full basis and ZX basis on an IBM quantum processor (\texttt{ibm\_kingston}). We show \texttt{1stp} ($N=18$) and \texttt{9aw2} ($N=14$). The bitstrings for both choices of basis follow the little endian convention, i.e., higher qubit indices are more significant, and therefore each Pauli observable is calculated using Qiskit's EstimatorV2 Primitive~\cite{ibm_estimator_primitive_2025}.}\label{table: measured Pauli strings}
\end{table*}

\section{MVWCP Resulting Graphs}
\label{sec: MVWCP graphs}

In Fig.~\ref{suppfig: weighted clique}, the ground-truth MVWCP solutions are provided for the two docking instances. The highlighted vertex sets correspond to mutually compatible pharmacophore contacts forming the optimal clique. Visualization of the clique structure confirms that the recovered solutions correspond to chemically interpretable contact patterns rather than isolated vertex selections, supporting the physical validity of the graph formulation.

\section{Additional Cost Functions as Evaluation Metrics for the Comparison of FBE and ZX-basis Encoding}
\label{sec: additional FBE and ZX comparison}

\begin{figure}[h]
    \centering
    \includegraphics[width=0.8\columnwidth,draft=False]{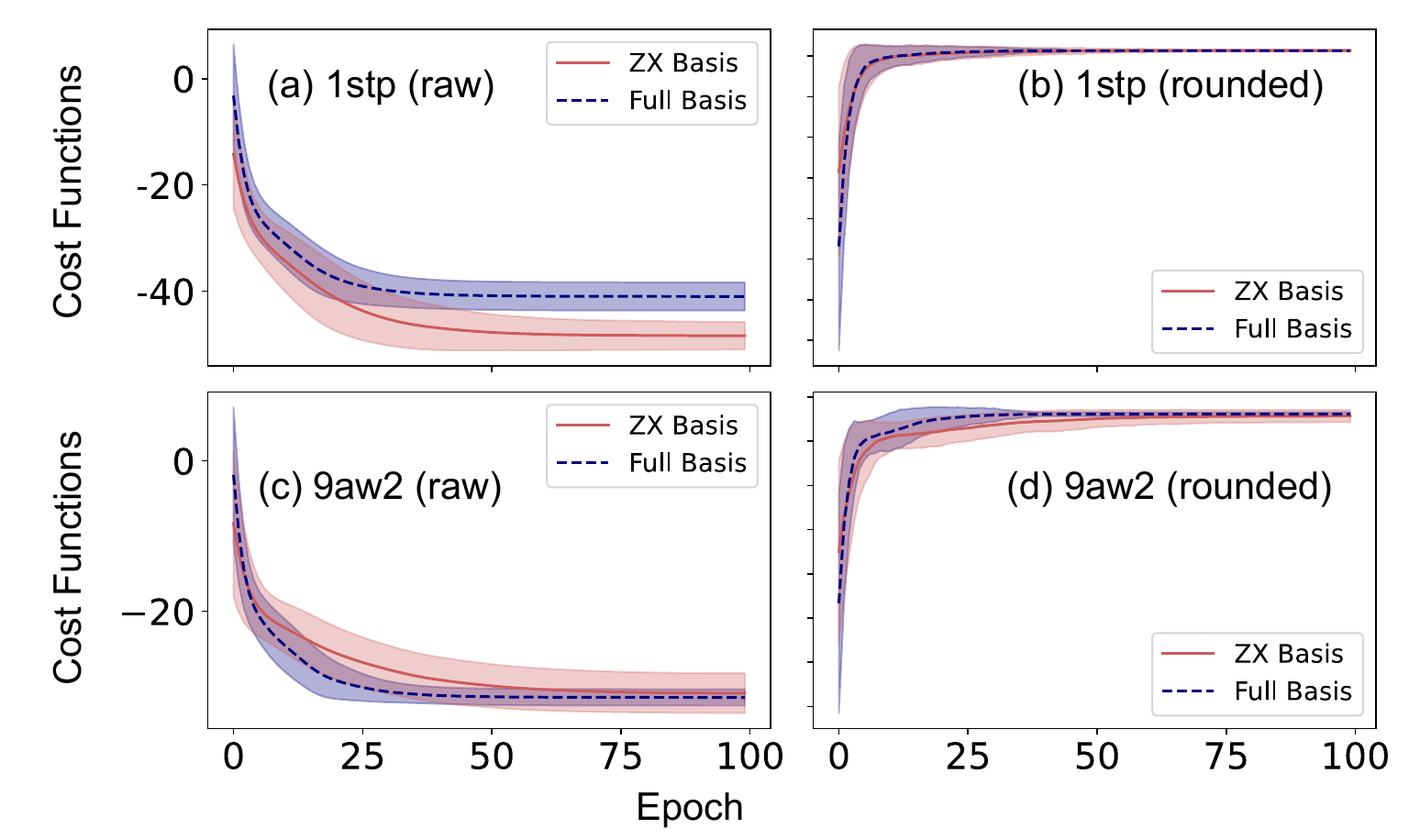}
    \caption{Comparison of raw and rounded cost function between the full basis encoding (blue dashed lines) and $ZX$ basis encoding (red solid lines). (a) raw cost function and (b) rounded cost function for \texttt{1stp}, and (c) raw cost function and (d) rounded cost function for \texttt{9aw2}. For all panels, the shaded region represents the standard error. Other parameters used are that the total epoch for training is kept at $100$.}
    \label{suppfig:other_functions_ZX_and_full_basis}
\end{figure}

In the main text, we report the transient clique weight during optimization. This quantity provides an interpretable physical measure corresponding to the weight of the candidate docking configuration selected at each training epoch. Comparison against the exact clique weight computed using classical graph solvers enables quantitative assessment of convergence reliability. It is found that FBE has an advantage over the ZX-basis encoding with early convergence during the optimization. In addition to the transient clique weight, in Fig.~\ref{suppfig:other_functions_ZX_and_full_basis}, for completeness, we also provide two more additional cost functions as the evaluation metrics, and compare optimization dynamics between full-basis and ZX encodings at the level of relaxed and rounded cost functions. Again, full-basis encoding consistently reduces both cost functions more rapidly and exhibits smaller variance across repetitions, similar to the transient clique weight behavior. These results suggest that our encoding does increase representational capacity during the optimization.

\begin{algorithm}[H]
  \caption{Warm-Start via Stochastic Imaginary Time Evolution for FBE}
\label{alg:warm_start_fbe}
\begin{algorithmic}[1]
\Require 
Hamiltonian $\hat{H} = \sum_{j=1}^{\mathcal{M}} \beta_j \hat{h}_j$; \\
Initial product state $|\psi_{+}\rangle = \bigotimes_{k=1}^M |+\rangle$; \\
Time step $\Delta t_{\text{WS}}$; \\
Number of Trotter steps $\mathcal{N}_{\text{T}}$.
\Ensure 
Warm-start state $|\psi_{\text{ini}}\rangle$ for FBE
\[
\mathbf{p}(\hat{h}_j) = \frac{|\beta_j|}{\sum_{j'=1}^{\mathcal{M}} |\beta_{j'}|}
\]

\State Initialize: $|\psi^{(0)}\rangle \gets |\psi_{+}\rangle$

\For{$t = 1$ to $\mathcal{N}_{\text{T}}$}
    \State Sample Hamiltonian term $\hat{h}_j$ according to $\mathbf{p}(\hat{h}_j)$
    \State Apply stochastic imaginary-time step (classically via MPS):
    \[
    |\psi^{(t)}\rangle \gets \frac{e^{-\Delta t_{\text{WS}}\hat{h}_j} |\psi^{(t-1)}\rangle}{\sqrt{\langle \psi^{(t-1)}| e^{-2\Delta t_{\text{WS}}\hat{h}_j} |\psi^{(t-1)}\rangle}}
    \]
\EndFor

\State Obtain warm-start state:
\[
|\psi_{\text{WS}}\rangle \gets |\psi^{(\mathcal{N}_{\text{T}})}\rangle
\]

\For{each qubit $i = 1$ to $M$}
    \State Measure local observables:
    \[
    \langle \sigma_i^x \rangle,\quad \langle \sigma_i^y \rangle,\quad \langle \sigma_i^z \rangle
    \]
    \State Compute rotation angles:
    \[
    \theta_i \gets \arccos(\langle \sigma_i^z \rangle)
    \]
    \[
    \phi_i \gets \arctan\left(\frac{\langle \sigma_i^y \rangle}{\langle \sigma_i^x \rangle}\right)
    \]
    \State Choose $\zeta_i$ arbitrarily
\EndFor

\State Construct warm-start quantum state:
\[
|\psi_{\text{ini}}\rangle \gets \bigotimes_{i=1}^M \left[ R_x(\phi_i) R_y(\theta_i) R_z(\zeta_i) |0\rangle_i \right]
\]
\State \Return $|\psi_{\text{ini}}\rangle$
\end{algorithmic}
\end{algorithm}

\end{document}